%% file: main.tex
\documentclass{article}
\usepackage[style=ieee,maxbibnames=3]{biblatex}
\usepackage{threeparttable}
\usepackage{subfigure}
\usepackage{graphicx}
\usepackage{MnSymbol}
\usepackage{adjustbox}
\usepackage{colortbl}
\usepackage{authblk}
\usepackage{changepage}
\usepackage{pdflscape}
\title{MRI-based and metabolomics-based age scores act synergetically for mortality prediction shown by multi-cohort federated learning}

\author[1]{Pedro Mateus*}
\author[2,3]{Swier Garst*}
\author[4,5]{Jing Yu*}

\author[2]{Davy Cats} 
\author[4]{Alexander G. J. Harms}
\author[4]{Mahlet Birhanu}

\author[2]{Marian Beekman} 
\author[2]{P. Eline Slagboom}
\author[3]{Marcel Reinders }
\author[12]{Jeroen van der Grond}

\author[1]{Andre Dekker} 
\author[6,7,8]{Jacobus F. A. Jansen} 
\author[9]{Magdalena Beran} 
\author[5,9]{Miranda T. Schram} 
\author[10]{Pieter Jelle Visser} 
\author[10,11]{Justine Moonen} 

\author[5]{Mohsen Ghanbari} 
\author[4,5]{Gennady Roshchupkin} 
\author[5]{Dina Vojinovic} 

\author[1]{Inigo Bermejo\textdagger}
\author[2]{Hailiang Mei\textdagger}
\author[4]{Esther E. Bron\textdagger}

\affil[1]{Department of Radiation Oncology (Maastro), GROW School for Oncology and Reproduction, Maastricht University Medical Centre\texttt{+}, Maastricht, the Netherlands.}
\affil[2]{Section of Molecular Epidemiology, Department of Biomedical Data Sciences, Leiden University Medical Center, the Netherlands}
\affil[3]{Delft Bioinformatics Lab, Delft University of Technology, Delft, the Netherlands}
\affil[4]{Biomedical Imaging Group Rotterdam, Department of Radiology \& Nuclear Medicine, Erasmus MC - University Medical Center Rotterdam, Rotterdam, the Netherlands}
\affil[5]{Department of Epidemiology, Erasmus MC - University Medical Center Rotterdam, Rotterdam, the Netherlands.}
\affil[6]{Department of Radiology and Nuclear Medicine, Maastricht University Medical Center, Maastricht, the Netherlands.}
\affil[7]{Mental Health \& Neuroscience Research Institute, Maastricht University, Maastricht, the Netherlands.}
\affil[8]{Department of Electrical Engineering, Eindhoven University of Technology, Eindhoven, the Netherlands.}
\affil[9]{Department of Internal Medicine, School for Cardiovascular Diseases (CARIM), Maastricht University, Maastricht, the Netherlands.}
\affil[10]{Alzheimer Center Amsterdam, Neurology, Vrije Universiteit Amsterdam, Amsterdam UMC location VUmc, Amsterdam, the Netherlands.}
\affil[11]{Amsterdam Neuroscience, Neurodegeneration, Amsterdam, The Netherlands.}
\affil[12] {Department of Radiology, Leiden University Medical Center, Leiden, the Netherlands.}

\begin{document}

\maketitle

\newpage

\begin{abstract}

\noindent Biological age scores are an emerging tool to characterize aging by estimating chronological age based on physiological biomarkers. Various scores have shown associations with aging-related outcomes. This study assessed the relation between an age score based on brain MRI images (BrainAge) and an age score based on metabolomic biomarkers (MetaboAge). We trained a federated deep learning model to estimate BrainAge in three cohorts. The federated BrainAge model yielded significantly lower error for age prediction across the cohorts than locally trained models. Harmonizing the age interval between cohorts further improved BrainAge accuracy. Subsequently, we compared BrainAge with MetaboAge using federated association and survival analyses.
The results showed a small association between BrainAge and MetaboAge as well as a higher predictive value for the time to mortality of both scores combined than for the individual scores. Hence, our study suggests that both aging scores capture different aspects of the aging process.

\end{abstract}

\input{introduction}

\input{results}

\input{conclusion_discussion}

\input{methods}

\printbibliography

\input{additional_info}

\setcounter{section}{0}
\setcounter{table}{0}
\setcounter{figure}{0}
\setcounter{subfigure}{0}
\input{supplementary_materials}
\end{document}

%% file: introduction.tex
\section{Introduction}

Understanding health in the context of aging is challenging, as aging encompasses various functional and structural changes in the body, including alterations in brain structure \cite{Vinke2018} and body metabolism \cite{deelen2019metabolic}. As people age, heterogeneity among individuals increases as some individuals may have larger health changes than what is common for their age. As a result, chronological age becomes less indicative of health in individuals at an older age. To address this issue, previous research introduced the concept of biological age, employing biomarkers based on physiological measurements \cite{Jylhava2017} \cite{Oh2023} \cite{Rutledge2022}. Such a biological age score may help to understand health in the context of aging and can provide a reference for identifying pathological changes.
\\\\

The biological age estimation methods consist of regression models optimized to predict chronological age from biomarker values in healthy aging individuals. These scores have been proposed based on various biomarkers using vastly different data modalities, reflecting different components of aging that can progress at a different pace between individuals. In the field of metabolomics, several biological age scores have been proposed based on blood-based metabolomics \cite{VanDenAkker2020} \cite{Hertel2015} \cite{Robinson2020} (e.g., MetaboAge). MetaboAge has been associated with cardiometabolic related outcomes, such as diabetes and heart failure, as well as more general aging-related phenotypes, such as decline in instrumental activities of daily living and all-cause mortality \cite{VanDenAkker2020}. In contrast to MetaboAge, MetaboHealth was trained to predict time to (all-cause) mortality instead of age. A higher MetaboHealth score indicates an increased chance of death within the next five years. In the field of neuroimaging, brain structure quantified with magnetic resonance imaging (MRI) was used to identify biological age predictors \cite{Franke2019} \cite{Cole2017} \cite{Wang2019} \cite{Jonsson} \cite{Bashyam} (i.e., BrainAge). BrainAge has been shown to predict mortality \cite{Cole2018}, various age-related diseases - such as dementia \cite{Wang2019}, Alzheimer's Disease (AD) and schizophrenia \cite{Bashyam}, and diabetes type 2 \cite{Franke2013} - and non-aging related diseases such as HIV \cite{Cole2017}.
While the predictive value of single brain-based and metabolomics-based age scores has been well studied, the relationships between these different biological age scores are largely unknown. Gaining a better understanding of this relationship may give insight on their added value and on how to optimally combine them to improve their predictive value. 
\\\\

For studying biological age scores, it is crucial that the regression models are trained in such a way that they generalize well to unseen data from other sources. Therefore, large-scale data from multiple studies and institutes is required \cite{Huisman2022}. The exact amount of data needed depends on the variability of the data in the eventual application as well as the complexity of regression models at hand. Current studies have proposed to use different types of regression models including conventional regression models like linear regression (i.e., \cite{VanDenAkker2020}) and large machine learning models based on deep neural networks (i.e., \cite{Wang2019}). To train the latter, which may include convolutional neural networks (CNNs) \cite{lecun95convolutional} that exploit complex patterns in the high-dimensional imaging data, access to large-scale and diverse training data is of utmost importance. The volume and diversity of data necessary for training is not usually owned by a single institution and, therefore, multi-centre collaborations are essential. However, privacy and safety concerns make it difficult, often impossible, to centrally collect data from multiple centres and make it available to train these models.
\\\\

In recent years, federated learning \cite{McMahan2017} has emerged as an approach to use sensitive data to train machine learning models while protecting privacy. Rather than training the model in a single institution (known as centralized learning), federated learning works by separately training at each institution's local computing nodes and only transferring aggregate statistics, like model parameters, between locations. A central server initiates the model parameters, aggregates the parameters sent back from each node after one or multiple epochs of local training on their local data, and then sends the aggregated parameters to each node. This routine is repeated until the model converges. As a result, it produces an optimized global model with knowledge of diverse local studies, which is trained over several distinct data collections without exchanging the data.
\\\\

Here, we apply federated learning to study the relationship between the two biological age scores, MetaboAge based on metabolites \cite{VanDenAkker2020} and BrainAge based on brain MRI \cite{Wang2019}, over three separate population-based cohorts. The main contributions include 1) insight into the relation between two biological age scores, 2) a federated learning infrastructure connecting three separate population-based cohorts, and 3) a comparison between federated and locally trained models.

%% file: results.tex
\section{Results}

\subsection{Study overview \textcolor{red}}

Three population-based cohorts  were included in a federated learning setup: the Rotterdam Study (RS) \cite{ikram2024Rotterdam}, The Maastricht Study (TMS) \cite{schram2014maastricht} and the Leiden Longevity Study (LLS) \cite{schoenmaker2006evidence}. The included number of subjects per cohort and their characteristics are shown in Table \ref{tab:characteristics}. For RS, 5,409 participants were included from the RS-I, RS-II and RS-III cohorts (2,509 participants had both MRI scans and blood samples available with taking lag time \textless{} 7 years) . For TMS and LLS, 5,055 (2,419) participants and 362 (362) participants were included, respectively. No demographic bias was found between participants with MRI scans only and participants with both MRI scans and blood samples available within each cohort. However, there are some biases between the cohorts. TMS has younger participants, while the RS has elder participants and a larger proportion of females. In addition, TMS has a much larger proportion of diabetes cases than the other two cohorts. 

\begin{table}[!htb]
\centering
\caption{Population characteristics summary of 3 cohorts included in the study}
\label{tab:characteristics}
\begin{threeparttable}
\resizebox{\textwidth}{!}{
\begin{tabular}{lccc} 
\hline
~ & \begin{tabular}[c]{@{}c@{}}The Rotterdam\\Study (RS)\end{tabular} & \begin{tabular}[c]{@{}c@{}}The Maastricht\\Study (TMS)\end{tabular} & \begin{tabular}[c]{@{}c@{}}The Leiden Longevity\\ Study (LLS)\end{tabular}  \\ 
\hline
participants with MRI scans ONLY\tnote{1}   & 2,900  & 2,636 & 0   \\

\hspace{2em}female, n (\%) & 1,583 (54.6)  & 1,331 (50.5) & -                                        \\

\hspace{2em}age at MRI, mean (sd)  & 67.0 (10.5)   & 59.6 (9.0)  & -                                      \\
participants with MRI scans AND blood samples   & 2,509    & 2,419 & 362                                                         \\

\hspace{2em}female, n (\%)           & 1,433 (57.1)      & 1,187 (49.0)  & 190 (52.5)                              \\
\hspace{2em}age at MRI, mean (sd)    & 67.5 (9.5)      & 60.3 (8.4)       & 65.5 (6.6)                                           \\

\begin{tabular}[c]{@{}l@{}} participants with complete covariates\\
eligible for association analysis  \end{tabular}    & 2,415  & 2,377    & 295     \\
\begin{tabular}[c]{@{}l@{}} \hspace{2em}lag time (years) between \\ \hspace{2em}blood sample and MRI scan, mean (sd)\end{tabular} & -1.39 (3.22) & 2.2 (1.3)   & 0                          \\

% \hspace{2em}mean age at blood sample (sd)                                                             & 67.0 (7.2)                                             & 59.3 (8.3)                                                & 59.3 (6.7)                                                       \\

% \hspace{2em}\% with BMI at blood sample avaliable (n)
%     & 100 (2509)                                          & 100 (2419)                                          & 99.5 (362)                                                              \\
\hspace{2em}BMI, mean (sd)   &   27.3 (3.8)                    & 26.6 (4.2)                      & 25.3 (3.3)  

                        \\
% \hspace{2em}\% with educational level avaliable (n)
%     & 99.5 (3145)                                         & 98.4 (2380)                                          & 100 (364)                                                               \\
    
\hspace{2em}educational level (low/median/high), \%
    & 46.3/29.7/24.0                       & 30.4/28.9/40.7                      & 55.2/8.6/36.2  
                        \\

% \hspace{2em}\% with type 2 diabetes diagnosis avaliable (n)
%             & 100 (2509)              & 100 (2419)             & 100 (364)                                     \\
\hspace{2em}diagnosis of diabetes, n (\%)
            & 221 (8.8)                                         & 538 (22.2)                                          &21 (5.8)                                                          \\
% participants eligible for survival analysis for dementia\
%             &2,407                                          & -                                             &294                                                              \\
\hspace{2em}diagnosis of dementia\tnote{2} , n (\%)\
            & 154 (6.4)                                         & -                                          
              & 3 (1.0)                                                             \\
\hspace{2em}follow-up time (years) of dementia, mean (sd)\
            &   6.9 (2.9)                                       & -                                          
              & 13.0  (2.5)                                                       \\

% participants eligible for survival analysis for mortality\
%             & 2,415                                         & -                                          
%               & 294                                                            \\
\hspace{2em}mortality, n (\%)\
            & 662 (27.4)                                         & -                                          
              & 48 (16.3)                                                              \\

\hspace{2em}follow-up time (years) of mortality, mean (sd)\
            &  10.0 (2.9)                                        & -                                          
              & 13.0 (2.5)                                                             \\

\hline
\end{tabular}
}

\begin{tablenotes}
\footnotesize 
\item[1]
Additionally includes participants with blood sampling and MRI scanning lag time 
\textgreater{} \\ 7 years, and excludes scans with dementia diagnose or stroke. 
\item[2]
8 missing values for dementia in RS and 1 missing value in LLS.

\end{tablenotes}

\end{threeparttable}
\end{table}

\noindent Based on these cohorts, we first trained the BrainAge model \cite{Wang2019} in a federated setting on the samples with only MRI data (RS and TMS) (i.e. the training and validation set) and evaluated the performance on the samples from all three cohorts with both MRI scans and blood samples (i.e. the test set), see Section \ref{sec:results_brainage}. Second, MetaboAge was computed for the test set samples using the original trained model in a previous study \cite{VanDenAkker2020} (Section \ref{sec:results_metabo}). We did not retrain the MetaboAge model since, in contrast to BrainAge, the original model was based on significantly more data from a large set of cohorts (18,000 samples from 26 cohorts). 
 Subsequently, we performed an association analysis between BrainAge and MetaboAge (Section \ref{sec:results_association}). Finally, we performed a survival analysis to study the complementary of BrainAge and MetaboAge for the association with time to mortality and dementia (Section \ref{sec:results_surv}).
 
\subsection{BrainAge and MetaboAge models\label{sec:results_brainage}}

\subsubsection{BrainAge model}
As shown in Table \ref{tab:BA_MAE}, the federated BrainAge model demonstrated its capability to predict the chronological age across different cohorts with mean absolute errors (MAE) of 5.59 years in the RS test set, 4.36 years in the TMS test set, and 4.60 years in the external test set (LLS). This federated model showed a better performance than local models that were tested on data from a different cohort. To highlight, the local model trained on RS data (training MAE = 2.48, test MAE = 4.21) achieved lower performance on TMS (test MAE = 6.45) and LLS (test MAE = 5.66) than the federated model. In addition, the local model trained on TMS data (training MAE = 3.10, test MAE = 4.72) also achieved lower performance on RS (test MAE= 7.29) and LLS (test MAE = 5.63) than the federated model, suggesting the two local models were unable to maintain performance when tested with data from a different cohort. Furthermore, the federated model showed more similar results of each subset compared to the local models, indicating less overfitting. These observations are supported by the results obtained with the 3-fold cross-validation approach (Supplementary Table \ref{tab:ba3fold}), which indicates similar MAE estimates for the three models.

\begin{table}[!htb] 
\centering
\caption{ MAE [95\% confidence interval] of BrainAge models trained locally and in a federated way. Models were trained and tested in The Maastricht Study (TMS) and the Rotterdam Study (RS), using the Leiden Longevity Study (LLS) as the external test cohort.}
\label{tab:BA_MAE}
\begin{threeparttable}
\begin{tabular}{llccc}  
\hline
~ & ~ & \multicolumn{2}{c}{Local models~} & Federated model  \\ 
\hline
  &  & TMS & RS & TMS \& RS \\ 
\hline
 TMS & Training & 3.10 [2.99, 3.23] & - & 4.75 [4.61, 4.90] \\ 
TMS & Validation & 4.59 [4.30, 4.87] & - & 5.56 [5.23, 5.89] \\ 
TMS & Testing & 4.72 [4.59, 4.87] & 6.45 [6.22, 6.71] & 5.59 [5.44, 5.76] \\ 
TMS & Testing \tnote{*}  & 4.67 [4.58, 4.73] & 6.55 [6.39, 6.71] & 4.97 [4.84, 5.12] \\
\hline
RS & Training & - & 2.48 [2.42, 2.54] & 4.34 [4.25, 4.43] \\ 
RS & Validation & - & 2.50 [2.37, 2.62] & 4.87 [4.66, 5.06] \\ 
RS & Testing & 7.29 [7.11, 7.44] & 4.21 [4.09, 4.34] & 4.36 [4.21, 4.48] \\ 
RS & Testing \tnote{*}  & 7.00 [6.22, 7.62] & 5.10 [4.95, 5.25] & 4.87 [4.74, 5.01] \\ 
\hline
LLS & Testing & 5.63 [5.24, 6.05]  & 5.66 [5.26, 6.08] & 4.60 [4.25, 4.95] \\ 
LLS & Testing \tnote{*}  & 5.82 [5.10, 6.25]& 5.89 [5.51, 6.30] & 4.21 [3.85, 4.60] \\

\hline
\end{tabular}

\begin{tablenotes}
\footnotesize 
\item[*]
Model trained with a sub-selection of the training participants with age between 53 and 75 years
\end{tablenotes}
\end{threeparttable}
\end{table}

\newpage
\noindent In addition, the federated BrainAge model results showed a predominantly higher MAE in TMS cohort when compared to the other cohorts. When training the federated BrainAge model with a sub-selection of the participants in the training set with age range between 53 and 75 years in both RS and TMS (70\% of the training set, mean age of 61.8 years), we observed that the model converged to a solution with smaller MAE differences (TMS test MAE of 4.97 vs 4.87 in RS and 4.21 in LLS).

\begin{figure}[!htb]
\centering
\subfigure[]{
\includegraphics[width=0.4\columnwidth]{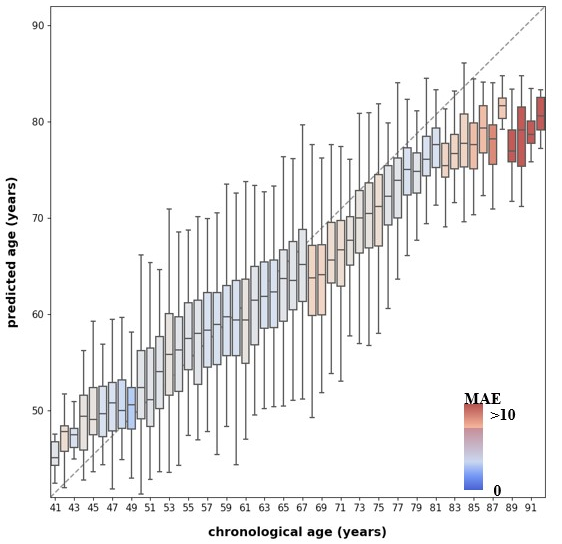} \label{fig:ba_boxplots}
}
\subfigure[]{
\includegraphics[width=0.4\columnwidth]{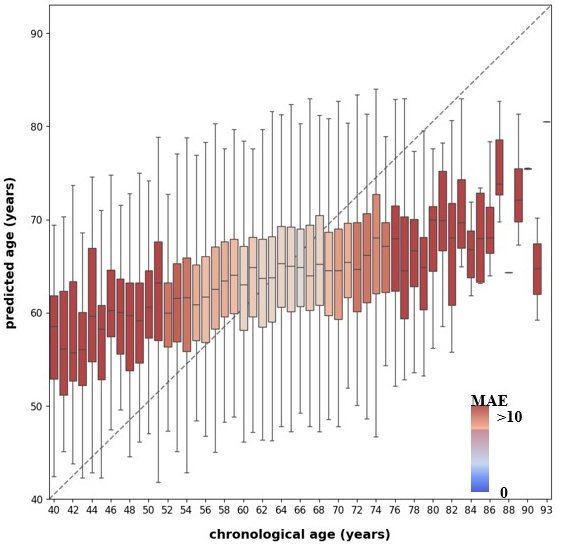}     \label{fig:metaboAges}

}
\caption{ Predicted age vs. chronological age for (a) Federated BrainAge and (b) MetaboAge, for the test sets of the three cohorts. The plot displays the distribution of the predicted age and associated Mean Absolute Error (MAE) by chronological age. X axis: the chronological age. Y axis: the predicted age. Color: the predicting MAE of participants at certain chronological age}

\end{figure}

\noindent We observed a tendency of the model to overpredict the age of younger subjects and underpredict the age of older subjects, as shown in Figure \ref{fig:ba_boxplots}. Although this tendency was apparent in all cohorts, differences existed in the age interval where it occurred between the cohorts (Supplementary Figure \ref{tab:bacohorts}). Applying a bias correction (See Methods) to the federated BrainAge model (See Supplementary Table \ref{tab:balinearcorrection}) resulted in considerable improvements for the RS and LLS but little for the TMS. Moreover, evaluating the bias correction with data from a single cohort, with either the TMS or the RS training set, displayed considerable improvements in the corresponding cohort but did not benefit external cohorts.

\subsubsection{Model optimization}

Model optimization for BrainAge is detailed in Supplementary Table \ref{tab:baoptimization}. In the federated architecture, selecting the model with the lowest MAE from each cohort showed optimal convergence. Performance was most similar between cohorts when models were aggregated between cohorts with equal importance weighting. Finally, increasing the number of epochs per round of training did not improve the performance.
\\\\
Optimized model hyperparameters were an initial learning rate of \(1 \times 10^{-3}\), a learning rate decay of \(1 \times 10^{-2}\), and a dropout rate in the last layer of \(5 \times 10^{-1}\). We employed the Adam algorithm to train the network for 20 rounds, three epochs each round and used a batch size of eight, the maximum possible due to computing memory limits. Computation time for federated training of the BrainAge model was 64\% higher on average than that of a central training (Table \ref{tab:Computation time}). 

\begin{table}[!htb]
\centering
\caption{ Computation time (range) taken to train the BrainAge model centrally and in a federated way (same computational resources in both approaches). The results presented comprise the time from the 3-fold cross-validation training.}
\label{tab:Computation time}

\begin{tabular}{ c c c c c }
\hline
 Trained & Data & \begin{tabular}[c]{@{}c@{}} Computation Time \\(hours)\end{tabular} & \begin{tabular}[c]{@{}c@{}} Number of \\Epochs\end{tabular} & \begin{tabular}[c]{@{}c@{}} Time / Epoch \\(min)\end{tabular}  \\
 \hline
 Centrally & TMS & 13.70 (12.93, 14.53) & 100 & 8.22 \\
 Centrally & RS & 16.83 (16.35, 17.23) & 100 & 10.10 \\
 Federated & TMS \& RS & 15.22 (13.23, 16.56) & 60 & 15.22 \\
 \hline
\end{tabular}
\end{table}

\subsubsection{MetaboAge model} 
\label{sec:results_metabo}
We applied the trained MetaboAge model from \cite{VanDenAkker2020} to obtain MetaboAge estimates in all three cohorts. Results can be seen in Figure \ref{fig:metaboAges}. This shows a larger age bias compared to BrainAge, which suffers even more from regression to the mean, as well as a wider range of predictions for each age bracket.

\subsection{The relation between BrainAge and MetaboAge }\label{sec:results_association}
To determine the relationship between the two biological age scores, federated linear regression analyses were used with MetaboAge as the dependent variable and BrainAge as the predictor. To correct for confounders, we built several models with a different number of confounding variables including age, sex, diabetes mellitus (DM), lag time, body mass index (BMI) and education category (EC). Continuous variables were normalized in the association analysis. Results are shown in Table \ref{tab:MA_betas}. A more detailed table including p-values and standard errors can be found in Supplementary Table \ref{tab:sup_betas_MA}.
\\\\
\noindent First, we considered a model with BrainAge as the only predictor (M1). This resulted in a small but significant association between BrainAge and MetaboAge (beta = 0.16, SE = 0.014, P = $4.3 * 10^{-32}$). Then, adding age as a covariate (M2) showed a strong effect on the observed relationship between BrainAge and MetaboAge (beta = -0.08, SE = 0.022, P = $6 * 10^{-5}$), indicating that the information both scores provide is chronological age. This effect of age on the relation between the two age scores was consistent when including sex, DM and lag time (M3-M4), as well as when further adding BMI and education (M5-M6). Adding other covariates apart from age didn't show strong influence on the relation between BrainAge and MetaboAge.
\begin{table}[!htb]
    \centering
    \caption{Beta values for various levels of covariates for estimating MetaboAge (a) and MetaboHealth (b).}
    \begin{threeparttable}
    (a) \\
    \resizebox{\textwidth}{!}{
    \begin{tabular}{cccccccccc}
        \hline
         Model ID & BrainAge & Age & Sex & DM$^1$ & Lag Time & BMI & EC1$^2$  & EC3$^2$ & Error (MAE)\\
         \hline
         M1 & 0.16$^*$$^*$ & - & - & - & - & - & - & - &   0.77\\
         M2 &-0.08$^*$ & 0.32$^*$$^*$ & - & - & - & - & - & - &   0.75 \\
         M3 & 0.25$^*$$^*$ & - & -0.17$^*$$^*$ & 0.16$^*$ &  0.08$^*$ & - & - & -  & 0.73\\
         M4 & -0.01$^=$ & 0.39$^*$$^*$ & -0.14$^*$$^*$ & 0.19$^*$ & 0.14$^*$$^*$ & - & - & - &   0.74\\
         M5 & 0.22$^*$$^*$ & - & -0.27$^*$$^*$& 0.08$^*$ & 0.01$^=$ & -0.03$^*$ & 0.16$^*$ &  0.11$^*$ & 0.76\\
         M6 & -0.06$^*$ & 0.38$^*$$^*$ & -0.23$^*$$^*$ & 0.08$^*$ & 0.04$^*$ & -0.03$^*$ & 0.07 $^*$ & 0.11$^*$ & 0.74\\
        \hline
    \end{tabular}
    }
    % *: P <0.05
    % =: P > 0.05
    \\
    (b) \\
    \resizebox{\textwidth}{!}{
    \begin{tabular}{cccccccccc}
        \hline
         Model ID & BrainAge & Age & Sex & DM$^1$ & Lag Time & BMI & EC1$^2$  & EC3$^2$ & Error (MAE)\\
         \hline
         M1 & 0.13$^*$$^*$ & - & - & - & - & - & - & - & 0.76\\
         M2 & 0.11$^*$& 0.03$^=$ & - & - & - & - & - & - &   0.76 \\
         M3 & 0.13$^*$$^*$& - & -0.04$^*$ & 0.70$^*$$^*$ & 0.02$^=$ &  - & - & - & 0.76\\
         M4 & 0.10$^*$& 0.02$^=$ & -0.04$^*$ & 0.71$^*$$^*$& 0.04$^*$ & - & - & - &   0.75\\
         M5 & 0.10$^*$$^*$ & - & -0.01$^=$ & 0.68$^*$$^*$ & 0.00$^=$& 0.15$^*$$^*$  & -0.07$^*$ &  -0.26$^*$$^*$ & 0.75\\
         M6 & 0.09$^*$& 0.06$^*$ & 0.06$^*$ & 0.66$^*$$^*$ & 0.03$^*$ & 0.10$^*$$^*$ & -0.10$^*$ & -0.27$^*$$^*$ & 0.74\\
        \hline
    \end{tabular}
    }
    \begin{tablenotes}
    \footnotesize 
    \item[1]
     DM = Diabetes Mellitus, i.e. diabetes (type 1 or 2) diagnosis. 
    \item[2]
     EC1-3 = Education Category, mapped to low/medium/high based on years of education. \\One-hot encoded relative to the medium level.
     \item[*]
     $P \leq 0.05$
     \item[**]
     $P \leq 5*10^{-10}$
     \item[=]
     $P > 0.05$
    \end{tablenotes}
    \end{threeparttable}
    \label{tab:MA_betas}
\end{table}

\noindent We also compared BrainAge with another metabolomics-based biomarker, MetaboHealth, which was estimated in all three cohorts applying the trained model from \cite{deelen2019metabolic}. Results can be found in Table \ref{tab:MA_betas} (b). Without any other covariates, the association between BrainAge and MetaboHealth was similar to that of MetaboAge. However, what differs was that this correlation persisted even after adjusted for age, suggesting that BrainAge and MetaboHealth share common information beyond chronological age.

\subsubsection{Survival analysis} 

\begin{landscape}
\label{sec:results_surv}
\begin{figure}[!htb]
    \begin{adjustwidth}{0cm}{0cm}
    \centering
    \subfigure[]{
        \centering
        \includegraphics[width = 0.2\paperwidth]{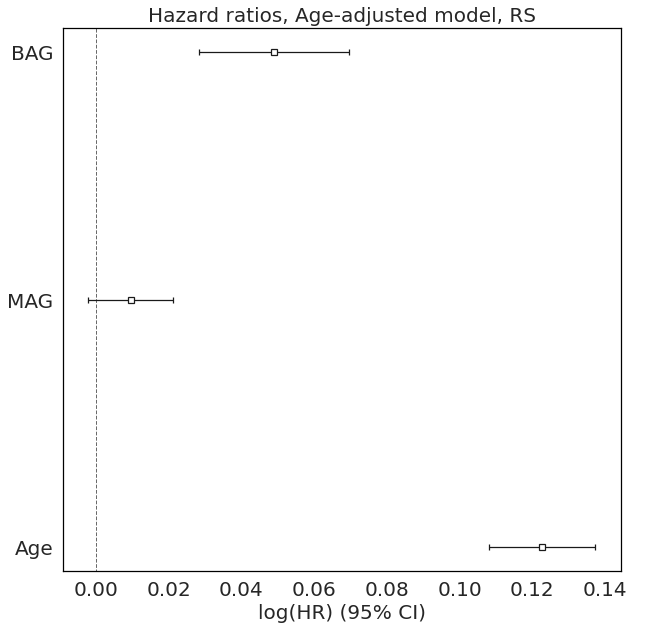}
    }
    \subfigure[]{
        \centering
        \includegraphics[width = 0.21\paperwidth]{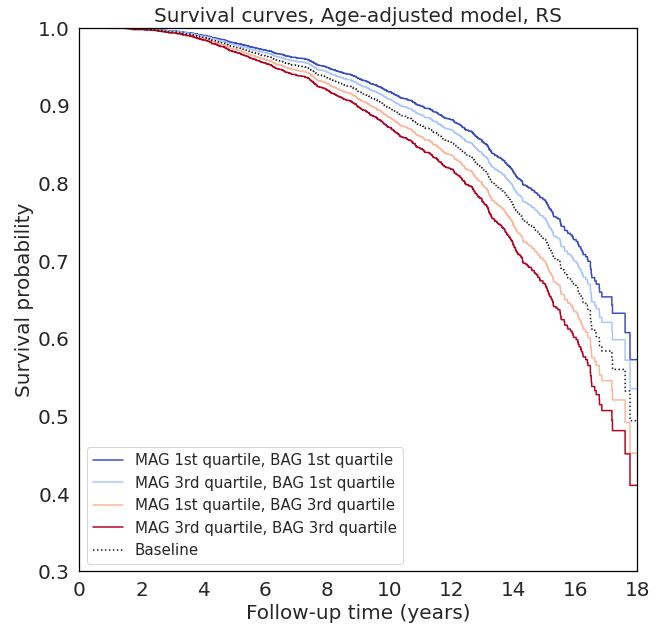}
    }  
    \subfigure[]{
    \centering
    \includegraphics[width = 0.2\paperwidth]{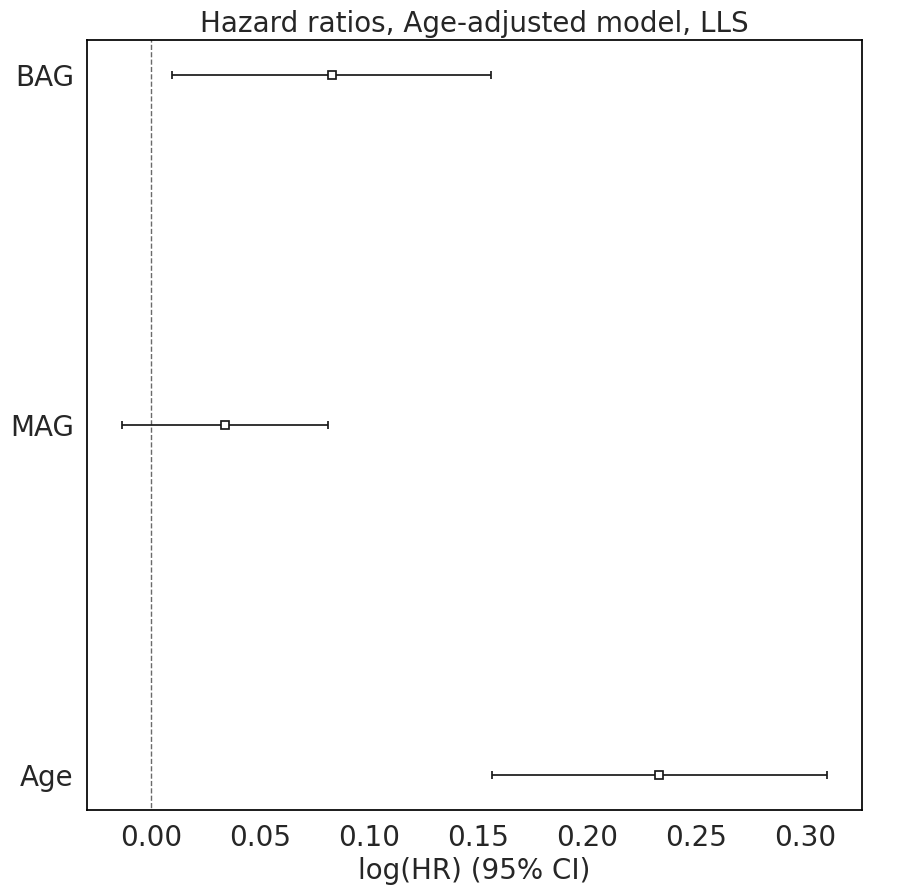}
    }
    \subfigure[]{
        \centering
        \includegraphics[width = 0.21\paperwidth]{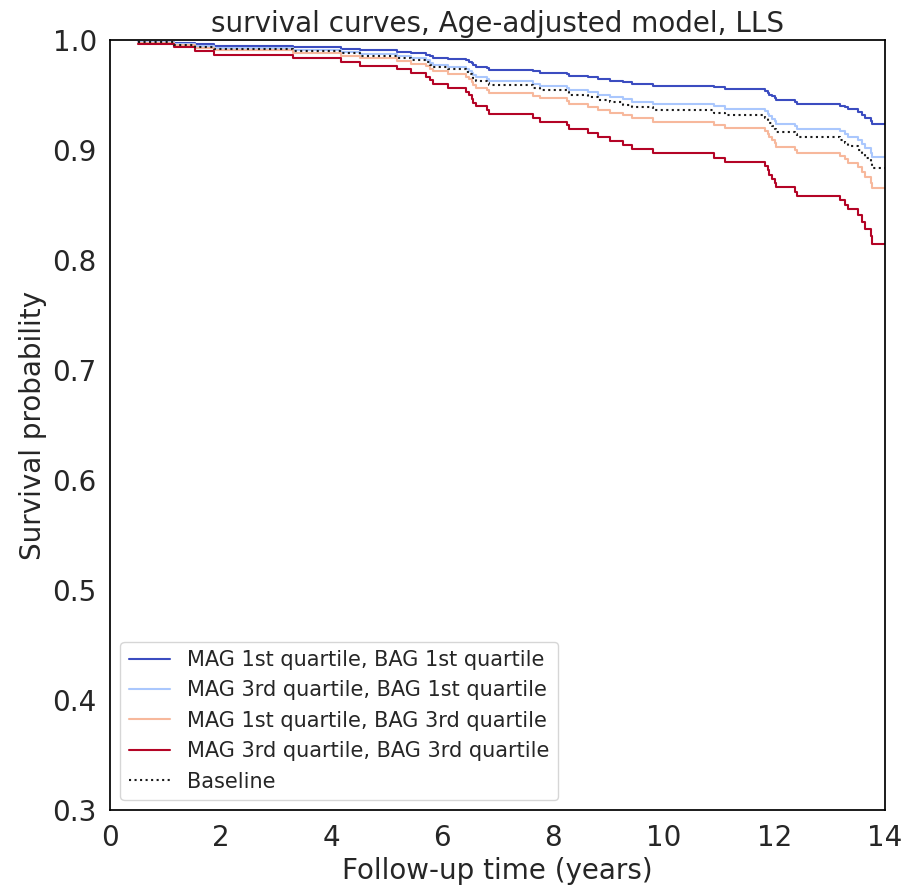}
    }
    \\
    
    \subfigure[]{
        \hspace{-11pt}
        \centering
        \includegraphics[width = 0.22\paperwidth]{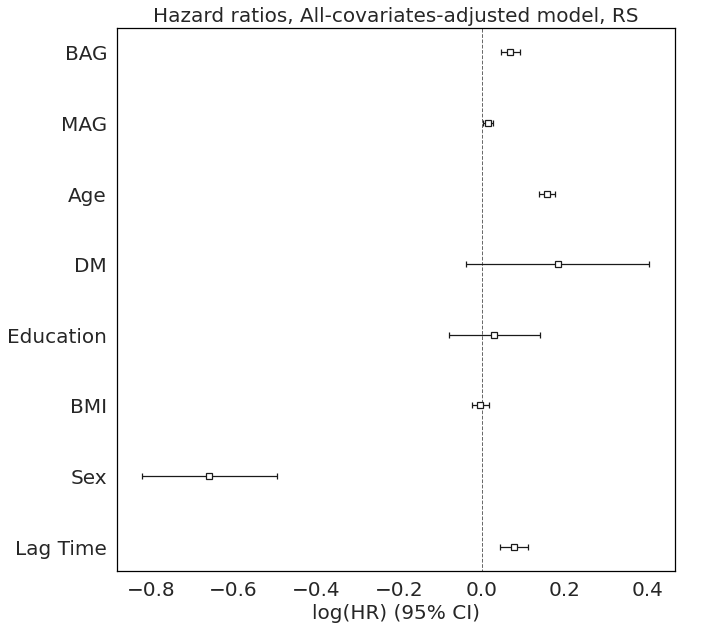}
    }
    \subfigure[]{
        \centering
        \includegraphics[width = 0.21\paperwidth]{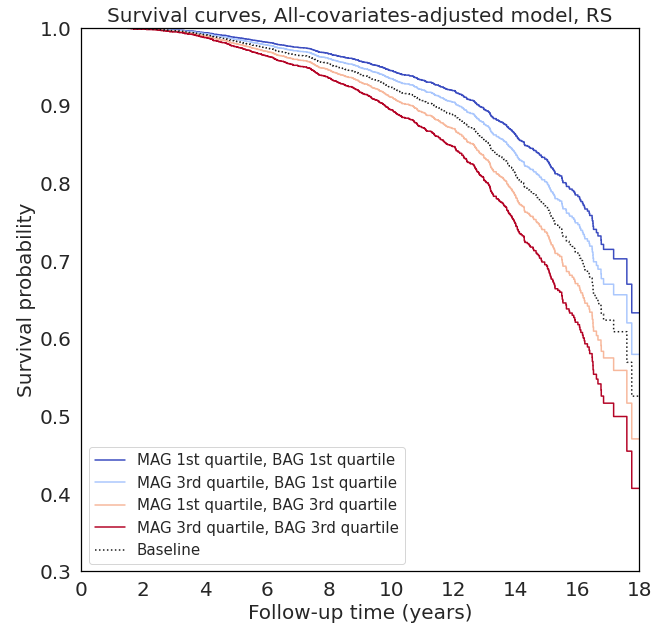}
    }  
    \subfigure[]{
    \hspace{-13pt}
    \centering
    \includegraphics[width = 0.22\paperwidth]{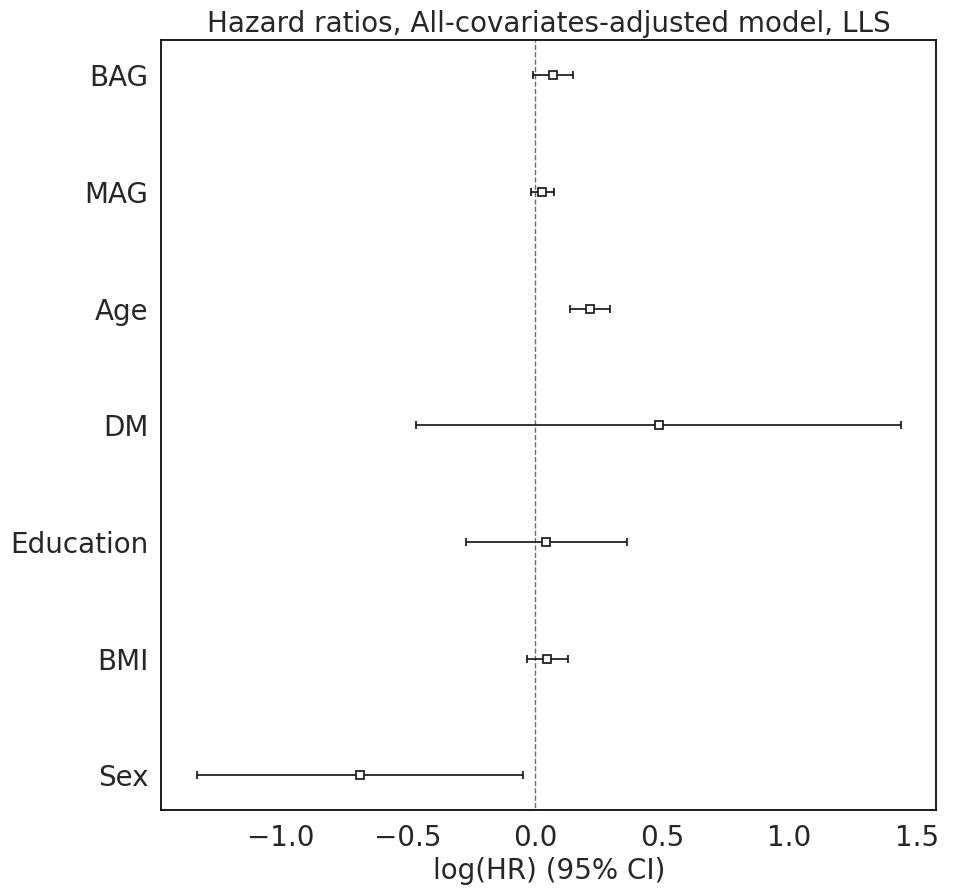}
    }
    \subfigure[]{
        \centering
        \includegraphics[width = 0.21\paperwidth]{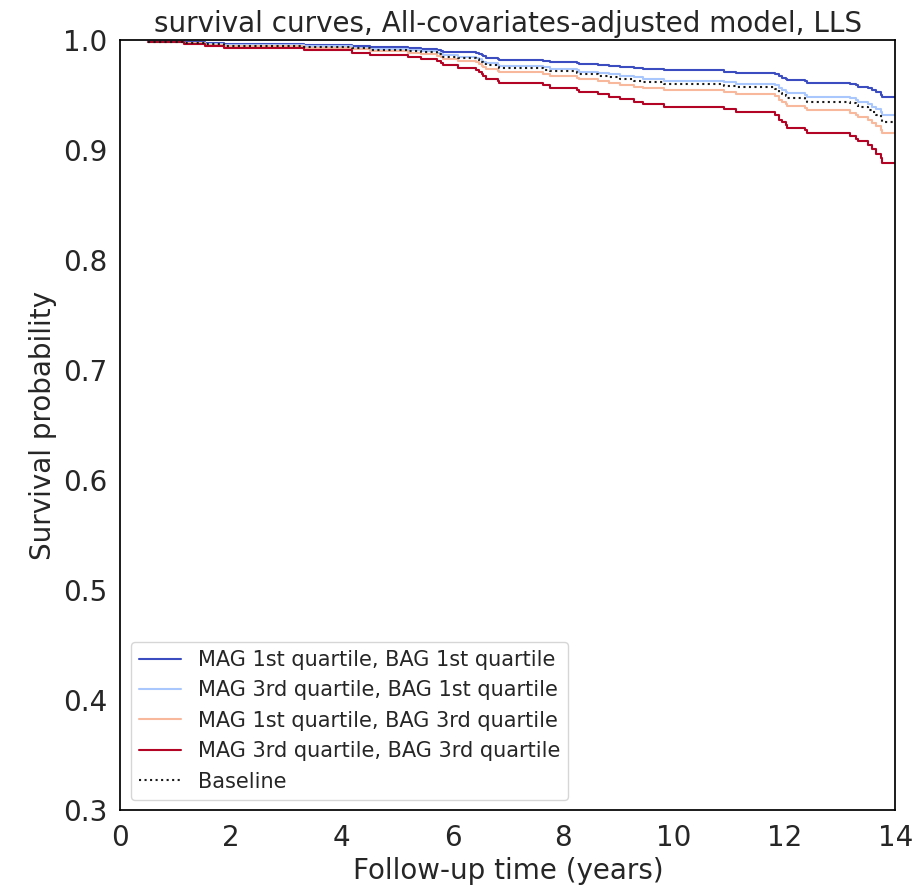}
    }
    \end{adjustwidth}
    \vfill
    \caption{Survival analysis for mortality prediction in the Rotterdam Study (RS) and the Leiden Longevity Study (LLS) using Cox Proportional Hazard models. Figure a-d show the results of the age-adjusted models (a-b in RS, c-d in LLS), taking only BrainAge Gap (BAG), MetaboAge Gap (MAG) and age into account. Figures e-h show the results for the all covariates-adjusted models (e-f in RS, g-h in LLS), additionally adjusting for Diabetes diagnosis (DM), education, Body Mass Index (BMI), sex and lag time. Note that the lag time in LLS is 0 for all participants. On the left are the hazard ratios of models, on the right the survival curves.}
    \label{fig:cph_mort}
\end{figure}

\end{landscape}

The weak (although significant) association between MetaboAge and BrainAge as presented in the previous section suggests that the two age scores carry different information. We therefore set out to see if we could combine both scores to increase the information about an individuals' health by performing the time to mortality and dementia prediction using survival analysis (see Methods). 
\\\\
We built Cox Proportional Hazards models based on BrainAge Gap (BAG: BrainAge - Age) and MetaboAge Gap (MAG: MetaboAge - Age), with age adjusted and all-covariates adjusted, on all participants. We then compared the survival curves of different subsection of the participants by taking the 1st and 3rd quartiles of both BAG and MAG. The 1$^{st}$ quartile represents a group with age scores lower than the chronological age (i.e. a relatively young appearing brain or metabolism), and the 3$^{rd}$ quartile represents the group of people with an age score higher than their chronological age (i.e. individuals with a relatively old-appearing brain or matabolism). We inspected the survival of different combinations of these two groups (1st-1st, 1st-3rd, 3rd-1st, and 3rd-3rd) using the estimated Cox Proportional Hazards models. Results of the survival analysis are shown in Figure \ref{fig:cph_mort}.  A clear separation between groups can be seen on survival curves. Young-appearing individuals on both biomarkers (MAG 1st quartile; BAG 1st quartile) showed the highest survival rate, while old-appearing individuals (MAG 3rd quartile, BAG 3rd quartile) showed the lowest survival rate. Indivuals scoring different on both markers had an intermediate survival rate. As suggested from the hazard ratios of BAG and MAG, this effect was more pronounced in RS than in LLS,  that both the BAG and MAG have more significant effects on the survival probability in RS. 
\\\\
When doing a simlar analysis to predict dementia, we found that only BAG was significantly associated with the time to dementia diagnosis, while MAG did not differentiate participants with dementia, independent of covariates (Supplementary Figure \ref{fig:cph_dementia}).

% \clearpage

%% file: conclusion_discussion.tex
\section{Discussion and Conclusion}

This study used federated learning to train and validate a BrainAge model across three cohorts and federated analysis to perform association and survival analysis of BrainAge and MetoboAge. Regarding BrainAge results, the performance of the federated BrainAge model was similar to other models reported in literature (MAE between 4-5 years) \cite{baecker2021machine}.
The federated model yielded significantly lower error (MAE) for
age prediction based on Brain MRI across cohorts than the locally trained models, showing that the federated model has better generalizability to external data. Such generalizability is a major concern for data methods in current medical practice \cite{steyerberg2016prediction}. Federated learning can enable cohorts with insufficient data to train an accurate model themselves to still get accurate model predictions. However, this is only possible if federated models generalize well to unseen cohorts.
\\\\
Age differences between cohorts had an impact on the results. We observed that the BrainAge model performance showed smaller MAE differences between cohorts when restricting participant selection to an equal age interval on all cohorts. Besides, we observed that both MetaboAge and BrainAge tend to overpredict the age of younger subjects and underpredict the age of older subjects, which is a known problem for biological age scores \cite{deLange2022} based on chronological age. We showed that a bias correction can help to decrease this tendency, however its effectiveness varies, especially for external test cohorts. 
\\\\
Regarding the federated association analysis, we found a low association between MetaboAge and BrainAge, which was drastically reduced after adjusting for age, indicating that the main association between the two biological age scores is their common correlation with age. In contrast, association between MetaboHealth and BrainAge did remain when adjusted for age. This could be related to earlier observations that MetaboHealth is (like BrainAge, and unlike MetaboAge) associated with cognitive decline \cite{Zonneveld2023}.
The low association between MetaboAge and BrainAge suggests complementarity of both scores, which is supported by the survival analyses.
\\\\
The survival analysis showed that individuals who scored high (indicating accerelated aging) on both BrainAge and MetaboAge had a lower survival rate than those that scored low on one or both of the scores. In the survival analysis results there are specific differences between cohorts that could be explained by study design differences between RS and LLS. Mortality was lower and BrainAge was less informative in LLS compared to RS. While LLS had inclusion criteria favouring healthy and long-living individuals, RS aims to include a general population and is less selective. We therefore hypothesize that LLS included participants with a relatively low BrainAge.
\\\\
One of the limitations of our analysis was the relatively limited amount of events in our survival analysis. Survival analysis on dementia was only possible in RS due to the lack of dementia cases in LLS. The amount of cases for mortality was 27.4\% and 16.3\% in RS and LLS, respectively. Performing a survival analysis with more cases could strengthen our results. Another limitation is the diversity of the populations. Although RS and TMS are population-based sudies, they mostly include participants  from western European descent, limiting the applicability of our findings to other populations. Finally, due to the inability to share data, our federated BrainAge model could not be compared to a centralized model trained on the same collection of data.
\\\\
Federated learning and federated analysis enable collaboration using data for which collaboration was not possible before, thereby increasing the pool of data for research. However, setting a federated infrastructure for real-life data also comes with several challenges. First, harmonized data pre-processing across cohorts is essential \cite{haddad2023multisite} \cite{doring2011evaluation}. We took account of this by harmonizing all data and reprocessing all imaging data with the same image analysis pipeline. Second, the distributed processing environment may provide challenges both for the optimization itself as well as for the optimization time needed. Regarding optimization, data heterogeneity between cohorts can lead to either overfitting in a single cohort or fluctuation in convergence between different cohorts when training the BrainAge model in a federated setting. By altering parameters such as lowering the local amount of epochs and increasing the dropout rate, we were able to decrease overfitting on the largest cohort ( see supplementary results \ref{sup:opt} for details). Furthermore, we experienced that the time needed for model optimization was relatively high as compute resource availability was not synchronized between cohorts. Third, data security should be taken into account properly. We used the Vantage6 implementation to address security issues using its user authentication system and the whitelisting option to moderate which algorithms are allowed to run. Fourth, Vantage6 was unable to interact directly with the cohort's high-performance compute platform required to train the BrainAge model. Therefore, we created a technical solution by extending the station node Docker image to allow establishing a connection to local HPC platforms. 
\\\\
In conclusion, this study demonstrated a federated BrainAge model that outperformed local models trained on only one cohort, highlighting that federated learning is a promising technique for cases in which data sharing is not possible. Our results additionally suggest that BrainAge and MetaboAge carry non-overlapping information with regard to time to all-cause mortality. We consider combining biological age scores based on different data modalities an interesting future research direction, as a combined age score will provide a more complete information for understanding health and may have a higher predictive value for identifying pathological changes in individuals.

%% file: methods.tex
\section{Methods}

\subsection{Data Preparation} 
\subsubsection{Study population}
We included participants from three cohort studies that take part in the Netherlands Consortium of Dementia Cohorts (NCDC): the Rotterdam Study (RS), the Maastricht Study (TMS), and the Leiden Longevity Study (LLS). The three cohort protocols include imaging and blood sample data necessary for our analysis. In addition, we used data from the Alzheimer’s Disease
Neuroimaging Initiative (ADNI) database (\url{adni.loni.usc.edu}) in the preparation of this article (detailed in the Supplementary Material).
\\\\
The Rotterdam Study is a prospective population-based study targeting causes and consequences of age-related diseases among 14,926 community-dwelling subjects aged 45 years and over \cite{ikram2024Rotterdam}. 
\\\\
The Maastricht Study is a prospective population-based study with focus on the etiology of type 2 diabetes of 10,000 individuals. It comprises individuals aged between 40 and 75 years from the southern regions of the Netherlands \cite{schram2014maastricht}. 
\\\\
The Leiden Longevity study includes 421 Caucasian families, each comprising long-lived siblings, along with their offspring and the spouses of the offspring. Families meeting the criteria for inclusion had a minimum of two long-lived siblings who were alive and willing to participate. Males were considered long-lived if 89 years or older and females 91 years or older \cite{schoenmaker2006evidence}.

\subsubsection{Data selection}

\begin{figure}[!htb]
    \centering
    \includegraphics[width=0.8\textwidth]{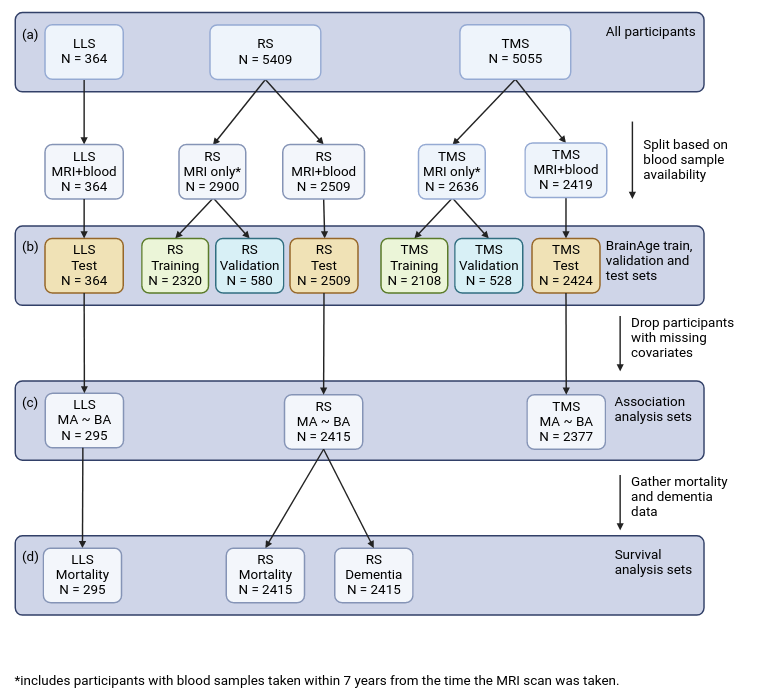}
    \caption{the data split flow. (a): The full studies considered in all three centers: Leiden Longevity Study ( LLS), Rotterdam Study (RS), and The Maastricht Study (TMS). (b): we created subsets based on which participants had blood draws taken, and split those in training, validation and test sets for the BrainAge training. (c): The participants that have both MRI scans and blood samples are used as test set for the BrainAge algorithm and are used for analysing the association of BrainAge (BA) with MetaboAge (MA). Covariates considered are: age, sex, lag time, BMI, diabetes diagnosis and education category.(d) We run the survival analysis on LLS and RS, since mortality and dementia data was unavailable at TMS. Dementia incidence was available at LLS, but was too low for performing a survival analysis (n = 3).}
    \label{fig:flowchart}
\end{figure}

For this analysis, the two main modalities used were T1-weighted MRI brain data \cite{Wang2019} and metabolomic data from the Nightingale metabolomics platform measured on blood draws \cite{VanDenAkker2020}. The data selection flowchart is shown in Figure \ref{fig:flowchart}. We included participants from the studies who had at least complete data of age, sex and brain MRI scans. For participants with metabolomic data, additional co-variates were (if available) diabetes mellitus, i.e. diabetes type 1 or 2 diagnosis, BMI and education level corresponding to their blood sampling time. In case blood samples and MRI scans were taken at different times (in RS and TMS), we used the interval years (lag time) between blood sampling time and MRI scanning time as an additional covariate.

\newpage
\noindent The data was first split into two parts, based on the availability of blood samples. Participants with blood samples and an absolute lag time smaller than seven years were used for testing the BrainAge model, the correlation analysis and the survival analysis. All other participants were randomly split into training (80\%) and validation (20\%) set for training the BrainAge model. The participants with missing values in the covariates were excluded from the association analysis. The demographic and clinical characteristics are in Table \ref{tab:characteristics}. We used all available scans for each subject in the longitudinal study. This allowed us to increase the number of training images, thereby introducing a natural type of data augmentation. Table \ref{tab:datasplits} presents the summary of data splits across cohorts. 

\begin{table}[!htb]
\centering
\caption{Data splits of 3 cohorts in the study} 
\label{tab:datasplits}
\begin{threeparttable}
\resizebox{\textwidth}{!}{
\begin{tabular}{lccc} 
\hline
~                     
    & \begin{tabular}[c]{@{}c@{}}The Rotterdam\\Study (RS)\tnote{1}\end{tabular} & \begin{tabular}[c]{@{}c@{}}The Maastricht\\Study (TMS)\end{tabular} & \begin{tabular}[c]{@{}c@{}}The Leiden Longevity\\ Study (LLS)\end{tabular} \\ 
\hline
Total number of scans            & 8318 & 5055 & 364 \\
Training and validation\tnote{2}        & 5809 & 2636 & - \\
Testing and correlation analysis & 2509 & 2419 & 364 \\ 
\hline
\end{tabular}
}
\begin{tablenotes}
\footnotesize 
\item[1]
Includes multiple scans at different time points for some participants
\item[2]
Includes healthy participants with MRI scans ONLY or with\\
blood draw and MRI scan lag time \textgreater{} 7 years
\end{tablenotes}
\end{threeparttable}
\end{table}

\subsubsection{Image acquisition}
Imaging data from the RS were obtained on a 1.5 T GE Signa Excite MRI scanner using an 8-channel head coil. The MRI protocol included a T1-weighted 3-dimensional (3D) Fast RF Spoiled Gradient Recalled Acquisition in Steady State with an inversion recovery pre-pulse (FASTSPGR-IR) sequence (TR = 13.8 ms, TE = 2.8 ms, TI = 400 ms, FOV = 25 × 25 cm2, matrix = 416 × 256 (interpolated to 512 × 512 resulting in voxel sizes of 0.49 × 0.49 mm2)\cite{de2009white}. 
\\\\
In TMS, the data were acquired on a 3T clinical magnetic resonance scanner (MAGNETOM Prismafit, Siemens Healthineers GmbH) using a head/neck coil with 64 elements for parallel imaging. The MRI protocol included a 3D T1-weighted magnetization prepared rapid acquisition gradient echo (MPRAGE) sequence (repetition time/inversion time/echo time (TR/TI/TE) 2300/900/2.98 ms, 176 slices, 256 × 240 matrix size, 1.0 mm cubic reconstructed voxel size) \cite{monereo2021quality}.
\\\\
Imaging in the LLS was performed on a Philips 3 Tesla Achieva MRI scanner using a standard 16-channel whole-head coil for radiofrequency transmission and reception (Philips Medical Systems, Best, The Netherlands). For each subject, a 3D T1-weighted anatomical scan was acquired with the following scan parameters: TR = 9.7 s; TE = 4.6 ms; flip angle = 8°; voxel size 0.88 x 0.88 x 1.40 mm.

\subsubsection{Image processing}
FreeSurfer version 6.0 \cite{fischl2012freesurfer} was used to segment supratentorial gray matter (GM) based on the T1-weighted brain MRI images\cite{fischl2004sequence}. GM density maps were computed based on an optimized voxel-based morphometry (VBM) protocol \cite{good2001voxel} \cite{roshchupkin2016fine} using the FSLVBM pipeline. First, all GM maps were nonlinearly registered to the standard Montreal Neurological Institute GM probability template (ICBM 152 Nonlinear atlases version 2009) with a 1 × 1 × 1 mm voxel resolution. Second, a spatial modulation procedure was used to avoid differences in absolute GM volume due to the registration. This is achieved by multiplying voxel density values by the Jacobian determinants estimated during registration. The matrix size of the modulated GM density maps was 196 x 232 x 188. As smoothing is a subgroup of possible mathematical operations which the network filters in the convolutional layer can represent, we did not apply smoothing on the VBM results. We performed a quality control based on the proportion (5\%) of outlier voxels and an additional manual check to exclude the outliers. Finally, we applied cropping and padding on the images to cut proper 0 edges, and masked the images with a k-Nearest-Neighbor-classifier segmented GM mask \cite{vrooman2007multi}. The matrix size of the final images was 160 x 192 x 144.

\subsection{Deep Learning for BrainAge prediction}

We used a 3D CNN model architecture proposed by \cite{Wang2019} to train a BrainAge model. This network takes as input the GM density maps obtained from the MRI scans and outputs a predicted age. The architecture consists of four convolutional blocks, used to extract valuable image features, followed by a fully connected layer that concatenates information on the participant's sex. We used the mean squared error (MSE) as the loss function to train the model and optimized the model parameters based on the model with the lowest MSE on the validation set. The model's accuracy was evaluated using the mean absolute error (MAE) on the test set. Both metrics measure the difference between model output and the participant's chronological age. To evaluate the associated uncertainty with each model, we performed bootstrapping with resampling (1000 resamples) to calculate the 95\% confidence interval. Additionally, to better estimate the model's performance, we performed a 3-fold cross-validation.

\subsubsection{Age-bias correction}

As observed in previous studies \cite{de2022mind}, BrainAge models are prone to overestimate the age of younger participants and underestimate the age of older participants. Since this behavior can impact subsequent analysis, an age-bias correction is normally applied using a linear regression model. In our study, we calculated three age-bias correction models based on \cite{de2020multimodal}'s approach, one for each training set separately (TMS and RS) and one for the federated approach. Additionally, we evaluated the generalizability of these models by assessing the performance in the cohorts' test sets.

\subsubsection{Federated Training}

We trained the BrainAge model using federated averaging (FedAvg) \cite{McMahan2017}. Initially, the deep learning model weights are randomly initialised and distributed to the participating cohorts. For every cohort, the model is individually trained for several epochs on their data, starting from the shared parameters. Next, the local model is sent back to the central server. Here, the model parameters are aggregated and shared with the training cohorts. This cycle continues until reaching the convergence criteria.

\subsubsection{Implementation Details}

We trained the deep learning model using Tensorflow \cite{abadi2016tensorflow} (version 2.8.0) and Python (version 3.8). The model's weights were initialized using the default Tensorflow method, the Xavier initialization \cite{glorot2010understanding}. Furthermore, we employed Docker \cite{Docker_2024} to containerize the scripts developed and provide the exact environment used in our experiments within a Docker image. A complete description of the libraries employed and the respective versions is provided in the public repository. Training and testing were performed at each cohort GPU cluster, specifically: an NVidia A40 GPU with 48GB and an NVidia RTX 2080 Ti GPU with 11GB for RS, a Tesla V100 with 32GB of RAM for TMS, and a TitanXp with 12GB for LLS. Finally, we followed the checklist for artificial intelligence in medical imaging (CLAIM) \cite{mongan2020checklist}, assessment provided in the Supplementary Material, to promote the reproducibility of our work.

\subsection{Metabolomics based age score (MetaboAge)}

We apply the trained model from Van den Akker et. al. \cite{VanDenAkker2020} to determine MetaboAge scores for the three cohorts. MetaboAge is a linear model based on a selection of 56 metabolites, as measured by the high-throughput proton nuclear magnetic resonance ($^1$H-NMR) metabolomics measurement platform Nightingale \cite{Wurtz2017}. The original model was trained and tested on a total of 18,716 blood samples, originating from 26 Dutch biobanks with ages ranging from 18 to 85 years. 

\subsection{Metabolomics based mortality score (MetaboHealth)}
MetaboHealth \cite{deelen2019metabolic} is a Cox proportional hazards model trained to predict all-cause mortality (contrary to BrainAge and MetaboAge, both trained to predict the age at measurement). MetaboHealth uses 14 metabolites from the high-throughput proton nuclear magnetic resonance ($^1$H-NMR) metabolomics measurement platform Nightingale \cite{Wurtz2017}, similar to MetaboAge. These were selected using a forward-backward process, that identified the metabolites with the lowest correlation with each other while being the most predictive for age at death. The original model was trained on 44.168 samples from 12 cohorts, with ages ranging from 18 to 110 years. Similar to MetaboAge, we apply this model on our cohorts to determine the MetaboHealth values used in this study.

\subsection{Vantage6 based Federated Learning Infrastructure}

\subsubsection{Vantage6 Personal Health Train system}
For the federated learning infrastructure, we adopted the Vantage6 Personal Health Train (PHT) framework \cite{moncada2020vantage6}. Vantage6 is a dockerized solution for federated learning, which comes with an access control system. A Vantage6 system consists of a central PHT server node and a set of distributed PHT station nodes. Each station node is located behind the institute's firewall and its control connection is regulated by the central server node, which uses a private key to determine which nodes can connect. Federated learning algorithms are implemented as Docker images and authorized by each station. 
\\\\
A federated training starts when a researcher sends a request to the PHT server. In each communication round, the PHT server sends a command to the participating station nodes with information on which Docker image to run. The stations execute this specific Docker image and send the results back to the central server. This server then saves these results in a database, from which the researcher can pull with their script.

\subsubsection{Job submission to local High Performance Cluster (HPC)}
Commonly, in a Vantage6 PHT system, all local jobs are executed at the station node which is often provisioned as a single virtual machine. However, this is insufficient for running the deep learning algorithms we need for BrainAge prediction with thousands of images. Therefore, we have created a technical solution by extending the station node Docker image (https://github.com/MaastrichtU-CDS/federated-brain-age/tree/master/v6\textunderscore wrapper) to be able to connect to each institute's local High Performance Cluster (HPC) facility such as a SLURM cluster (at LLS and RS) or an on-premises Kubernetes managed cluster (Data Science Research Infrastructure in TMS). When a task is submitted, instead of specifying the algorithm image, a placeholder Docker image name is specified with the actual algorithm image being specified in the task's inputs. In the station node configuration this placeholder is mapped to a locally available Docker image (wrapper image). This wrapper image will be executed by the station node to redirect the task to the local HPC, download and verify the federated learning algorithm image. A local file transfer command, such as \textit{scp}, is used inside the wrapper image to transfer the input data, task inputs and token to the cluster's file system. A job to run the algorithm image will be submitted to the cluster scheduler and this job's status will be monitored though \textit{ssh}. Once the job has finished, the output file produced by the algorithm image will be retrieved and written to the output file as expected by Vantage6.

\subsubsection{Security and privacy}
\begin{figure} %\label{fig:PHT_security}
    \centering
    \includegraphics[width=1\textwidth]{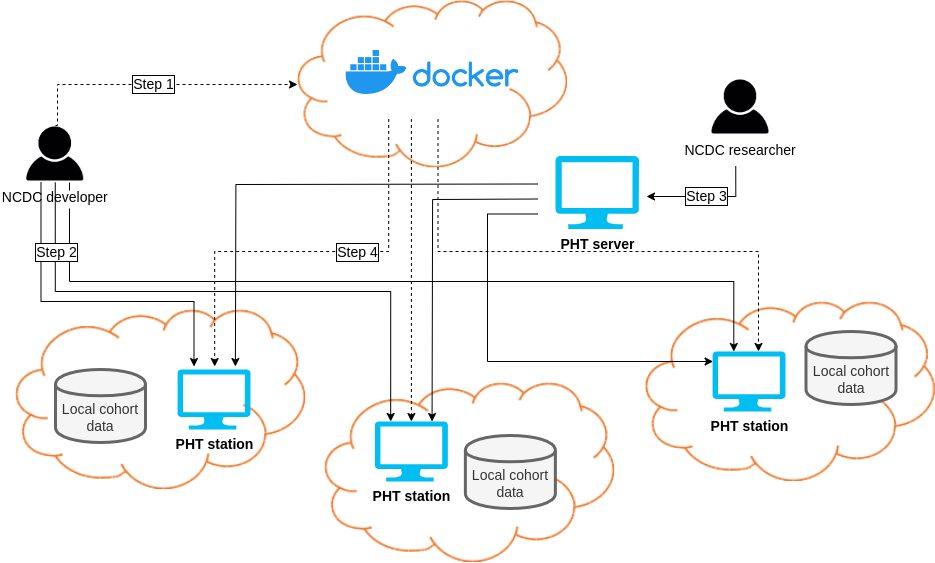}
    \caption{NCDC PHT governance protocol to ensure security and data privacy requirements. Step 1: NCDC developers implement and verify the PHT algorithms (in Docker image). SHA256 digests of Docker images are generated. Step 2: NCDC developers support cohort owners to examine these algorithms (e.g., which datasets are retrieved and transferred), then together decide which algorithms are allowed in their local PHT station (using allowed\textunderscore images tag). Step 3: NCDC researcher performs data analysis via a PHT server. Step 4: Only authorized PHT algorithms are executed at each cohort's local PHT station.}
    \label{fig:pht_security}
\end{figure}

Data security and privacy preservation is a core requirement in our analysis as defined in the project agreement so that no sensitive data (e.g., brain MRI images) will be transferred outside of a cohort intentionally or unintentionally. We have established a governance protocol to address this. As illustrated in Figure \ref{fig:pht_security}, first our project developers implement and jointly verify the Vantage6 PHT algorithm Docker image. Second, local institutional developers will support the cohort owners to examine these algorithm Docker images (e.g., which datasets are accessed and analyzed and which aggregated results are transferred to the server node). Third, the cohort owners decide whether to whitelist an algorithm image using its specific SHA256 digest to run on their local node (configured through the allowed\textunderscore images field in the node configuration). Any unauthorized changes in the PHT algorithm image will result in a different SHA256 digest and cause this algorithm to be denied at a PHT station. Therefore, from a cohort owner perspective, only verified and certified Vantage6 algorithms can access their local data to ensure the privacy. All data communication between PHT server and PHT stations are encrypted using RSA to further guarantee the data security.

\subsubsection{Data management}

Although Vantage6 addresses a wide range of requirements for a federated system, it does not provide an out-of-the-box solution to guarantee data interoperability between the station nodes. To address this, we harmonized the clinical data in each station node using a data model as described in \cite{dadata}. Each station includes a local PostgreSQL database connected to the PHT node that guarantees structural and semantic data compatibility. Additionally, we employed this database to store the deep learning models and performance metrics. Regarding the imaging data, we homogenized the storage systems across cohorts by using the open source Extensible Neuroimaging Archive Toolkit (XNAT \cite{marcus2007extensible} to store the MRI scans. If not available, a central XNAT server was available for use. By storing the MRI scans in equal storage systems, the same data structure was enforced across the cohorts. After this harmonization step, we transferred the necessary imaging data to the GPU cluster for training the deep learning models, avoiding a high throughput of read and write operations.

\clearpage

\subsection{The relation between BrainAge and MetaboAge }
\subsubsection{Association Analysis}
We explored the association between MetaboAge and BrainAge through linear regression using MetaboAge as the explanatory variable, and BrainAge as the response variable:  
\begin{equation}
    Y = \beta_b * BrainAge + \sum_{x_i \in X} \beta_i * x_i
\end{equation}
with $Y$ being MetaboAge, and with $X$ being the set of covariates. Three sets of covariates were considered:
\begin{enumerate}
    \item Adjustment for age:$X = \{Age\}$
    \item Additional adjustment for sex, DM (diabetes diagnosis) and lag time: \\ $X = \{Age, Sex, DM, Lag\ Time\}$
    \item Additional adjustment for BMI and education category (EC): \\ $X = \{Age, Sex, DM, Lag\ Time, BMI, EC1, EC3\}$
\end{enumerate}
Additionaly, we trained a model with no covariates. Since age seemed to have a large effect on these models, we ran both 2 and 3 with and without age, resulting in 6 sets of covariates in total. We ran these 6 models both for MetaboAge as well as for MetaboHealth as explanatory variable.
\\\\
This linear regression is also performed using the PHT infrastructure. Each round $t$, the PHT server sends out global beta values ($\beta_g = \{\beta_b, \beta_1,..,\beta_i\}$) to all cohorts. These then create a local update of the beta values ($\beta_l$) using one iteration of gradient descent:
\begin{equation}
    \beta_l^{t} = \beta_g^{t} - \eta * \Delta L
\end{equation}
with $L$ being the mean squared error loss function. $\eta$ was set at 0.1 for all models. Then, each cohort sends back their own $\beta_l$ to the PHT server, which creates a new value for $\beta_g$ using a weighted average:
\begin{equation}
    \beta_g^{t+1} = \frac{1}{\sum_{j=0}^J n^j} \sum_{j = 0}^J n^j * \beta^t_{l,j} 
\end{equation}
with $\beta^{t}_{l,j}$ being the local beta values coming from cohort $j$ at round $t$. This iterative process continued until MAE did not change anymore. We repeat this process 10 times, and choose the model with the lowest MAE. The beta values of this linear regression model are used for measuring the association of the involved covariates.

\subsubsection{Comparison to meta-analysis}
\noindent The association analysis was conducted using the federated approach. However, since we were running simple linear regressions without any regularization, there was a closed-form solution to find the optimal beta values for each covariate. Instead of federated analysis through PHT, this solution could also be efficiently calculated based on meta-analytical frameworks, such as HASE \cite{roshchupkin2016hase}. We therefore compared our results with the closed-form solution provided by HASE.

\subsubsection{Survival analysis}

To assess the complementary value of BrainAge and MetaboAge in estimating the vulnerability of individuals, we performed survival analyses using Cox proportional hazards models. As input to these survival analyses, we used the difference between participants' age score and their chronological age, MAG and BAG (MAG = MetaboAge - Age; BAG = BrainAge - Age). Using these gaps, we fit a Cox proportional hazards model:
\begin{equation} \label{eq:cox}
    \lambda(t|MAG, BAG, X) = \lambda_0(t) * exp(\beta_1*MAG + \beta_2*BAG + \sum_{x_i \in X} \beta_{x_i}x_i)
\end{equation}
With X being the set of covariates we adjusted for. We considered two sets: adjustment for age only ($X = \{Age\}$), as well as adjustment for the full set of covariates which were used in the association analysis \\
($X = \{Age, Sex, DM, Lag 
 time, BMI, EC_1, EC_3\}$). 
\\\\
We determined the cutoff point at the 1$^{st}$ and 3$^{rd}$ quartiles for both BAG and MAG. Intuitively, at this first cutoff point, the gap is lower than average, indicating that the biological age score is relatively low for a given age, indicating less aging than expected. Conversely, at the second cutoff, the age score is relatively higher than average after adjusting for age, which means that the score indicates accelerated aging. In total, this creates four groups:
\begin{itemize}
    \item BAG 1$^{st}$ quartile, MAG 1$^{st}$ quartile : little aging according to both BrainAge and MetaboAge
    \item BAG 1$^{st}$ quartile, MAG 3${rd}$ quartile : little aging according to BrainAge, accelerated aging according to MetaboAge
    \item BAG 3$^{rd}$ quartile, MAG 1$^{st}$ quartile : accelerated aging according to BrainAge, little aging according to MetaboAge
    \item BAG 3$^{rd}$ quartile, MAG 3$^{rd}$ quartile: accelerated aging according to both BrainAge and MetaboAge
\end{itemize}

\noindent We then applied the trained model from equation \ref{eq:cox} using the BAG and MAG values for these four groups, creating four different survival curves. The survival analyses were run locally in RS with mortality and dementia as outcomes, and in LLS on mortality only as this cohort had only few dementia cases at the latest follow-up (N = 3); no long-term mortality and/or dementia data was available from TMS.

%% file: additional_info.tex
\section*{Author contributions}

P.M., S.G., and J.Y. conducted the analysis and prepared the results. P.M., D.C., A.G.J.H., and M.Birhanu implemented the infrastructure. M.Beekman, P.E.S., J.G., J.J., M.Beran, M.T.S., M.G., G.R., and D.V. took part on the collection and curation of data. P.E.S., M.T.S., P.J.V., J.M., I.B., H.M, and E.E.B. coordinated the project and prepared the legal agreements. P.M., S.G., J.Y., I.B., H.M., and E.E.B. wrote the original draft. All authors provided respective domain knowledge expertise during the course of the research and edited the manuscript. P.M., S.G., and J.Y. have contributed equally to this work and are co-first-authors. I.B., H.M., and E.E.B. have jointly supervised this work.

\section*{Acknowledgements}

The authors acknowledge all researchers of the Netherlands Consortium of Dementia Cohorts (NCDC), funded in the context of Deltaplan Dementie from ZonMW Memorabel (projectnr 73305095005) and Alzheimer Nederland. Furthermore, this project has received funding from the VOILA Consortium (ZonMw grant nr. 457001001). J.Y. was sponsored by China Scholarship Council (No. 202006640010) from the Ministry of Education of P.R. China.
\\\\
This research was made possible, in part, using the Data Science Research Infrastructure (DSRI) hosted at Maastricht University.
\\\\
Data collection and sharing for this project was funded by the Alzheimer's Disease Neuroimaging Initiative (ADNI) (National Institutes of Health Grant U01 AG024904) and DOD ADNI (Department of Defense award number W81XWH-12-2-0012). ADNI is funded by the National Institute on Aging, the National Institute of Biomedical Imaging and Bioengineering, and through generous contributions from the following: AbbVie, Alzheimer’s Association; Alzheimer’s Drug Discovery Foundation; Araclon Biotech; BioClinica, Inc.; Biogen; Bristol-Myers Squibb Company; CereSpir, Inc.; Cogstate; Eisai Inc.; Elan Pharmaceuticals, Inc.; Eli Lilly and Company; EuroImmun; F. Hoffmann-La Roche Ltd and its affiliated company Genentech, Inc.; Fujirebio; GE
Healthcare; IXICO Ltd.; Janssen Alzheimer Immunotherapy Research \& Development, LLC.; Johnson \& Johnson Pharmaceutical Research \& Development LLC.; Lumosity; Lundbeck; Merck \& Co., Inc.; Meso Scale Diagnostics, LLC.; NeuroRx Research; Neurotrack Technologies; Novartis Pharmaceuticals Corporation; Pfizer Inc.; Piramal Imaging; Servier; Takeda Pharmaceutical Company; and Transition Therapeutics. The Canadian Institutes of Health Research is providing funds to support ADNI clinical sites
in Canada. Private sector contributions are facilitated by the Foundation for the National Institutes of Health (\url{www.fnih.org}). The grantee organization is the Northern California Institute for Research and Education,
and the study is coordinated by the Alzheimer's Therapeutic Research Institute at the University of Southern California. ADNI data are disseminated by the Laboratory for Neuro Imaging at the University of Southern California.

\section*{Competing interests}

The author(s) declare no competing interests.

\section*{Data Availability}

The data from the Leiden Longevity Study, The Maastricht Study and Rotterdam Study are accessible for researchers upon request. Please find the required forms at: \url{https://leidenlangleven.nl/data-access/}, \url{https://www.demaastrichtstudie.nl/research/data-guidelines}. Requests for Rotterdam Study data can be directed to data manager Frank J.A. van Rooij (\url{f.vanrooij@erasmusmc.nl}). 

\section*{Code Availability}

The underlying code for this study is available in GitHub and can be accessed via the following links: 
\begin{itemize}
\item Association analysis: \url{https://github.com/swiergarst/association_ analysis}.
\item Federated neural network for BrainAge: \url{https://github.com/MaastrichtU-CDS/federated-brain-age}.
\item Vantage6 node extension: \url{https://github.com/MaastrichtU-CDS/ncdc-memorabel}.
\end{itemize}

%% file: supplementary_materials.tex
\pagebreak
\section{Supplementary Results \label{sec:suppl_adni}}

\subsection{BrainAge }

\subsubsection{Federated vs centralized training}
\label{sec:ADNI}

To compare the performance of the BrainAge model between a federated and a centralized setting, we conducted an assessment with publicly available data. For this purpose, we simulated three cohorts and optimized the BrainAge model hyperparameters for the federated approach. Data used in the preparation of this supplement were obtained from the Alzheimer’s Disease Neuroimaging Initiative (ADNI) database (\url{adni.loni.usc.edu})\footnote{The investigators within the ADNI contributed to the design and implementation of ADNI and/or provided data but did not participate in analysis or writing of this report.
A complete listing of ADNI investigators can be found at:
\url{http://adni.loni.usc.edu/wp-content/uploads/how_to_apply/ADNI_Acknowledgement_List.pdf}}. The ADNI was launched in 2003 as a public-private partnership, led by Principal Investigator Michael W. Weiner, MD. The primary goal of ADNI has been to
test whether serial magnetic resonance imaging (MRI), positron emission tomography (PET), other biological markers, and clinical and neuropsychological assessment can be combined to measure the
progression of mild cognitive impairment (MCI) and early Alzheimer’s disease (AD).
\\\\
We randomly divided the data from healthy individuals into three groups (N=221, 219, 217) and placed it in each cohort's station, simulating a federated infrastructure with three separate institutes. MRI data was pre-processed following the pipeline described in the Methods section.
To evaluate the model's performance, we applied a leave-one-out cross-validation strategy by using one cohort exclusively for testing in each round. Additionally, we trained the model centrally with the complete dataset to assess the baseline performance.
\\\\
The results show that the federated model converged to a solution without requiring different hyperparameters from the central model (Table \ref{tab:adniresults}). The BrainAge estimation from the federated learning model (average testing set MAE of 4.13) reaches similar performance as centralized learning (testing set MAE of 3.94). 
\\\\
In conclusion, this experiment showed that the federated model performed similarly to the model trained centrally and did not benefit from changing the hyperparameters.

\begin{table}[!htb]
\centering
\caption{Comparative performance (MAE (range)) of the BrainAge model trained with ADNI data centrally or with the federated architecture. The results presented comprise the average MAE and range (values in brackets) from the 3-fold cross-validation.}
\label{tab:adniresults}
\begin{tabular}{ c c c c c }
\hline
& Training & Validation & Testing \\ 
\hline
Central model & 2.22 (2.00, 2.42)  & 3.91 (3.55, 4.29) & 3.94 (3.54, 4.34) \\
Federated model & 2.35 (1.66, 3.67)  & 4.39 (4.05, 4.73) & 4.13 (4.02, 4.35) \\
\hline
\end{tabular}
\end{table}

\subsubsection{Model optimization} \label{sup:opt}

The application of the federated BrainAge model to the NCDC cohorts revealed the need to further improve the hyperparameters and aggregation methods compared to the initial experiment with ADNI data. The heterogeneity of the dataset affected the model's convergence and exacerbated the problem of overfitting one of the cohorts.
\\\\
To assess the impact of the model training options (number of epochs by round, model selection, and weighted averaging based on sample size), we repeated a 3-fold cross-validation for each. The results in Table \ref{tab:baoptimization} suggest that a smaller number of epochs, no weighted averaging, and selecting the local model with higher MAE improves the performance. Regarding the hyperparameters, we observed that a higher learning rate decay (\(1 \times 10^{-2}\) vs \(1 \times 10^{-4}\)) and dropout rate (0.5 vs 0.25) benefited convergence.

\begin{table}
\centering
\caption{ Performance evaluation (MAE (range)) for the federated training optimization (test set).}
\label{tab:baoptimization}
\resizebox{\textwidth}{!}{
\begin{tabular}{ c c c c c c c }
\hline
 \begin{tabular}[c]{@{}c@{}}Weighted \\ averaging \tnote{1}\end{tabular} & \begin{tabular}[c]{@{}c@{}}Model \\ selection \tnote{1}\end{tabular} & \begin{tabular}[c]{@{}c@{}}Number of \\ epochs \tnote{1}\end{tabular} & TMS & RS & LLS \\ 
\hline
$\checkmark$ & - & 3 & 6.71 (6.13, 7.17) & 4.18 (4.04, 4.39) & 5.62 (4.22, 6.85) \\ 
- & - & 3 & 7.04 (6.30, 7.73) & 4.24 (4.03, 4.44) & 4.67 (4.02, 5.21) \\ 
$\checkmark$ & $\checkmark$ & 3 & 7.53 (6.94, 8.41) & 4.67 (4.02, 5.77) & 4.54 (4.44, 4.67) \\ 
- & $\checkmark$ & 3 & 6.34 (5.29, 7.00) & 4.68 (4.53, 4.95) & 4.43 (4.00, 5.16) \\ 
- & $\checkmark$ & 6 & 7.92 (5.56, 9.73) & 4.22 (4.06, 4.32) & 4.47 (4.19, 4.70) \\ 
\hline
\end{tabular}
}
\end{table}

\subsubsection{Population differences}

\begin{figure}[h]
\centering
\resizebox{\textwidth}{!}{
\includegraphics[scale=0.7]
{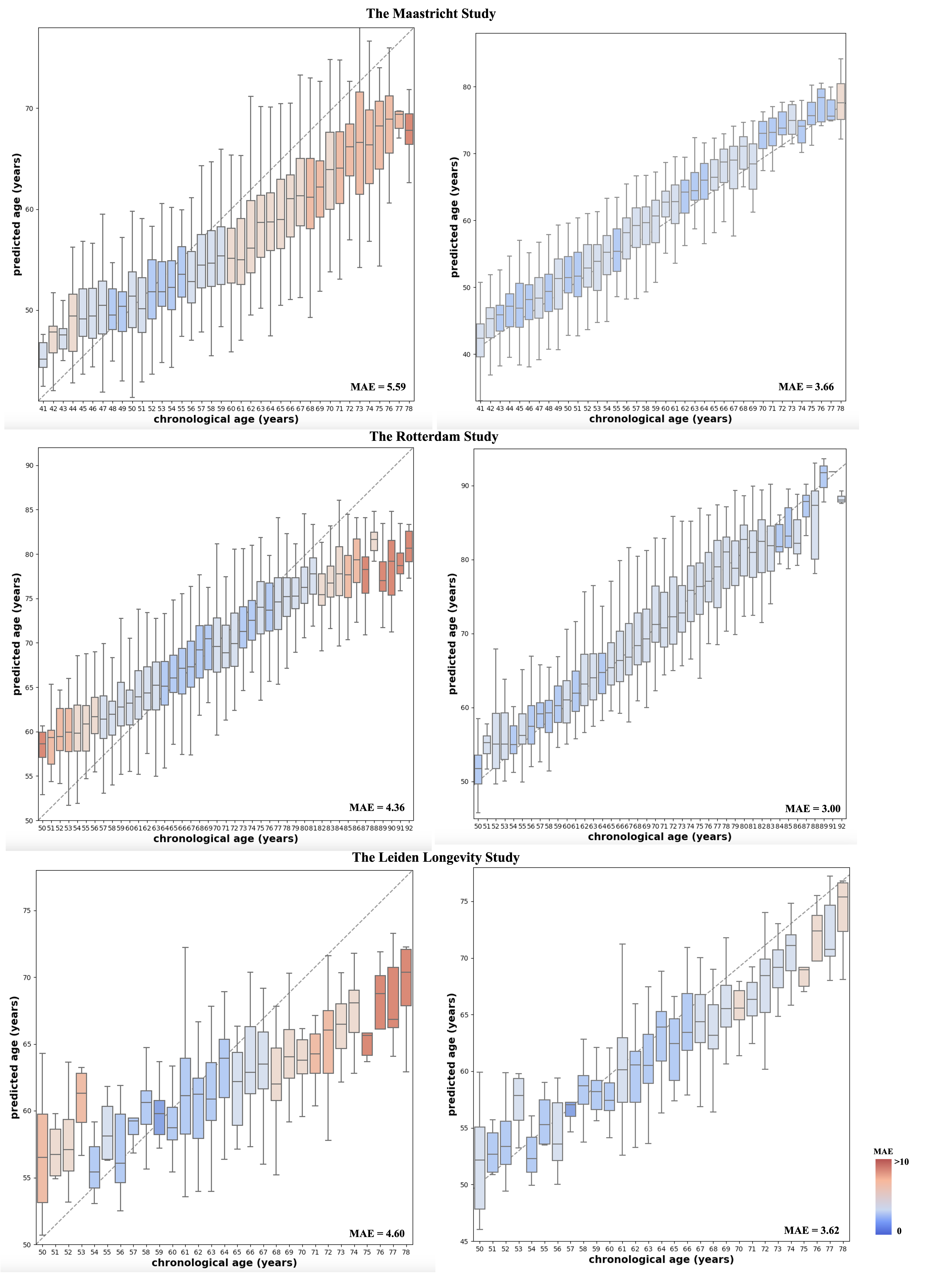}
}
\caption{Brain age prediction using the federated model (on the left) and applying the linear correction (on the right) for the test set of each cohort.}
\label{tab:bacohorts}
\end{figure}

The performance differences observed in the BrainAge models trained locally or in a federated collaboration hinted at the possible impact of the population differences. As shown in Table \ref{tab:ba3fold}, locally trained models display a higher MAE range in unseen data compared to the federated approach. Moreover, Figure \ref{tab:bacohorts} highlights this challenge by indicating different age intervals per cohort where the model underestimates and overestimates the chronological age. Consequently, applying a linear correction to the BrainAge predictions appears to be a cohort-specific solution.
\\\\
When applying a bias correction to the federated BrainAge model (Table \ref{tab:balinearcorrection}), it resulted in notable improvements for the RS and LLS (MAE of 3.33 and 3.62 vs 4.36 and 4.60) but little for the TMS (MAE of 5.51 vs 5.59). Moreover, evaluating the bias correction with data from a single cohort, with either the TMS or the RS training set, displayed considerable improvements in the corresponding cohort (MAE of 3.66 for TMS and 3.00 for RS) but did not benefit external cohorts (MAE of 6.31 for TMS and 7.41 for RS).

\begin{table}
\centering
\caption{ 3-fold cross-validation performance (MAE (range)) of BrainAge models trained locally and using the federated approach (test set).}
\label{tab:ba3fold}
\begin{tabular}{ c c c c c }
\hline
 & TMS & RS & Federated \\ 
\hline
TMS & 4.67 (4.58, 4.73) & 7.00 (6.22, 7.62) & 6.34 (5.29, 7.00) \\ 
RS & 7.88 (7.55, 8.46) & 4.18 (4.16, 4.22) & 4.68 (4.53, 4.95) \\ 
LLS & 5.82 (5.10, 6.25) & 5.23 (4.44, 6.32) & 4.43 (4.00, 5.16)  \\
\hline
\end{tabular}
\end{table}

\begin{table}
\centering
\caption{MAE of the federated BrainAge model trained with TMS and RS data when applying the age-bias correction models (test set). Three models were tested, one for each training set separately and one for the complete training set (TMS and RS). Values in brackets represent the 95\% confidence interval.}
\label{tab:balinearcorrection}
\resizebox{\textwidth}{!}{
\begin{tabular}{ c c c c c }
\hline
 & Without linear correction &  & With linear correction &  \\ 
 & - & TMS \& RS & TMS & RS \\ 
\hline
TMS & 5.59 [5.44, 5.76] & 5.51 [5.37, 5.67] & 3.66 [3.54, 3.77] & 7.41 [7.25, 7.58]  \\ 
RS & 4.36 [4.21, 4.48] & 3.33 [3.23, 3.44] & 6.31 [6.17, 6.45] & 3.00 [2.91, 3.08] \\ 
LLS & 4.60 [4.25, 4.95] & 3.62 [3.35, 3.88] & 3.41 [3.13, 3.70] & 4.73 [4.42, 5.03]  \\
\hline
\end{tabular}
}
\end{table}

\clearpage
\newpage

\subsection{The relation between BrainAge and MetaboAge}

\subsubsection{Full table of beta values including p-values and standard errors}

\begin{table}[!htb] 
\centering
\caption{Beta values (beta), p-values (P) and standard errors (SE) for various levels of covariates for estimating MetaboAge (a) and MetaboHealth (b).}
\label{tab:sup_betas_MA}
\begin{threeparttable}
(a) \\
\resizebox{\textwidth}{!}{
\begin{tabular}{ll|llllll}
\hline
            &      & M1             & M2      & M3     & M4      & M5 & M6 \\
            \hline
            & beta & 0.16           &  -0.08  &  0.25  &  -0.01  & 0.22  & -0.06   \\
BrainAge    & P    & 4.3*$10^{-32}$ &  6.0*$10^{-5}$  &  2.2*$10^{-66}$   & 3.1*$10^{-1}$   & 2.3*$10^{-49}$   &  3.7*$10^{-3}$  \\
            & SE   & 0.014          & 0.022 & 0.014   & 0.022   & 0.015   & 0.022   \\
            \hline
            & beta &                &  0.32   &    & 0.39  &    &  0.38   \\
Age         & P    &                & $4.5*10^{-50}$   &    & 1.8*$10^{-65}$   &    & 7.6*$10^{-63}$    \\
            & SE   &                & 0.022   &    & 0.022   &    &  0.022  \\
            \hline
            & beta &                &    & -0.16   & -0.14   & -0.27   & -0.23   \\
Sex         & P    &                &   & 1.2*$10^{-17}$    & 6.6*$10^{-14}$   & 1.5*$10^{-27}$   & 3.2*$10^{-21}$   \\
            & SE   &                &    &  0.019  & 0.019   & 0.025  & 0.024   \\
            \hline
            & beta &                &    &  0.16   & 0.19   & 0.08   &  0.08  \\
DM\tnote{1}           & P    &                &    & 3.25*$10^{-6}$   & 5.8*$10^{-08}$   & 2.1*$10^{-2}$   &  1.4*$10^{-2}$  \\
            & SE   &                &    &  0.036  & 0.035   & 0.038   & 0.037   \\
            \hline
            & beta &                &   & 0.08   &  0.14  & 0.01   &  0.04  \\
Lag Time    & P    &                &   & 2.5*$10^{-9}$   &  4.3*$10^{-22}$  & 2.1*$10^{-1}$   & 2.1 * $10^{-3}$   \\
            & SE   &                &    & 0.014 & 0.014  &  0.014 & 0.014   \\
            \hline
            & beta &                &   &    &    &  -0.03  & -0.03   \\
BMI         & P    &                &    &    &    & 1.5*$10^{-2}$  &  6.5*$10^{-3}$  \\
            & SE   &                &    &    &    & 0.014   & 0.014   \\
            \hline
            & beta &                &    &    &    & 0.16  & 0.07   \\
EC1\tnote{2}          & P    &                &    &    &    & 6.0*$10^{-9}$    & 4.4 * $10^{-3}$   \\
            & SE   &                &    &    &    &  0.028  &  0.028  \\
            \hline
            & beta &                &    &    &    & 0.11   & 0.11   \\
EC3\tnote{2}          & P    &                &    &    &    & 1.6*$10^{-5}$   & 1.1*$10^{-5}$   \\
            & SE   &                &    &    &    & 0.027   & 0.027   \\
            \hline
Error (MAE) &      &  0.77              & 0.75   & 0.73   & 0.74   & 0.76   & 0.74  \\
\hline
\end{tabular}
}
\\
(b) \\

\resizebox{\textwidth}{!}{
\begin{tabular}{ll|llllll}
\hline
            &      & M1             & M2      & M3     & M4      & M5 & M6 \\
            \hline
            & beta & 0.13           &  0.11  &  0.13  &  0.10  & 0.10  & 0.09   \\
BrainAge    & P    & 5.3*$10^{-20}$ &  1.7*$10^{-6}$  &  2.3*$10^{-20}$   & 2.1*$10^{-6}$   & 1.5*$10^{-13}$   &  1.4*$10^{-5}$  \\
            & SE   & 0.014          & 0.022 & 0.014   & 0.022   & 0.014   & 0.022   \\
            \hline
            & beta &                &  0.03   &    & 0.02  &    &  0.06   \\
Age         & P    &                & $9.3*10^{-2}$   &    & 1.8*$10^{-1}$   &    & 2.1*$10^{-3}$    \\
            & SE   &                & 0.022   &    & 0.022   &    &  0.022  \\
            \hline
            & beta &                &    & -0.04   & -0.04   & -0.01   & 0.06   \\
Sex         & P    &                &   & 8.6*$10^{-3}$    & 1.6*$10^{-2}$   & 2.9*$10^{-1}$   & 9.7*$10^{-3}$   \\
            & SE   &                &    &  0.019  & 0.019   & 0.024  & 0.024   \\
            \hline
            & beta &                &    &  0.7   & 0.7   & 0.68   &  0.66  \\
DM\tnote{1}          & P    &                &    & 9.07*$10^{-86}$   & 1.4*$10^{-86}$   & 1.9*$10^{-73}$   &  1.5*$10^{-68}$  \\
            & SE   &                &    &  0.035  & 0.035   & 0.037   & 0.037   \\
            \hline
            & beta &                &   & 0.02   &  0.03  & 0.00   &  0.03  \\
Lag Time    & P    &                &   & 6.6*$10^{-2}$   &  3.6*$10^{-1}$  & 2.1*$10^{-1}$   & 1.1 * $10^{-2}$   \\
            & SE   &                &    & 0.014 & 0.014  &  0.014 & 0.014   \\
            \hline
            & beta &                &   &    &    &  0.15  & 0.10   \\
BMI         & P    &                &    &    &    & 1.2*$10^{-26}$  &  1.5*$10^{-13}$  \\
            & SE   &                &    &    &    & 0.014   & 0.014   \\
            \hline
            & beta &                &    &    &    & -0.07  & -0.10   \\
EC1\tnote{2}        & P    &                &    &    &    & 7.4*$10^{-3}$    & 2.4 * $10^{-4}$   \\
            & SE   &                &    &    &    &  0.027  &  0.028  \\
            \hline
            & beta &                &    &    &    & -0.26   & -0.28   \\
EC3\tnote{2}         & P    &                &    &    &    & 9.1*$10^{-23}$   & 2.5*$10^{-26}$   \\
            & SE   &                &    &    &    & 0.026   & 0.026   \\
            \hline
Error (MAE) &      &  0.77              & 0.77   & 0.74   & 0.74   & 0.72   & 0.72  \\
\hline
\end{tabular}
}
\begin{tablenotes}
    \footnotesize 
    \item[1]
     DM = Diabetes Mellitus, i.e. diabetes (type 1 or 2) diagnosis. 
    \item[2]
     EC1-3 = Education Category, mapped to low/medium/high based on years of education. \\One-hot encoded relative to the medium level.
    \end{tablenotes}
\end{threeparttable}
\end{table}

\clearpage
\subsubsection{Comparison to meta-analytic framework}

We compared the MataboAge regression model calculated through our
iterative federated approach and those from a meta-analysis approach called
HASE [32] (see Methods). The comparison between federated linear regression MAE’s and HASE
MAE’s is in Table \ref{tab:HASE_comp}.

\begin{table}[!htb]
    \centering
    \caption{Comparison between federated linear regression MAE's and HASE MAE's}
    \begin{tabular}{ccc}
    \hline
         Model & federated MAE & HASE MAE\\
         \hline
        M1 & 0.77 & 0.77\\
        M2 & 0.75 & 0.75\\
        M3 & 0.74 & 0.73\\
        M4 & 0.74 & 0.73\\
        M5 & 0.76 & 0.76\\
        M6 & 0.76 & 0.76\\
        \hline
    \end{tabular}
    \label{tab:HASE_comp}
\end{table}

\begin{figure}[!htb]
    \centering
    \resizebox{\textwidth}{!}{
    \includegraphics[width=0.5\textwidth]{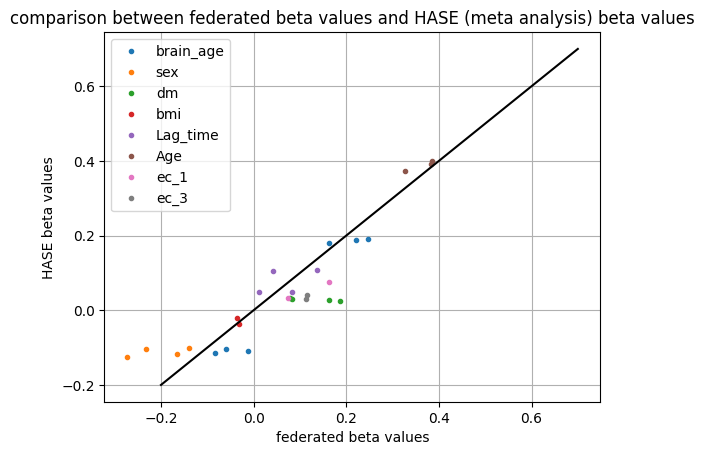}
    }
    \caption{Comparison of federated beta values with HASE beta values }
    \label{fig:hase_pht_comparison}
\end{figure}

\noindent Figure \ref{fig:hase_pht_comparison} presents a comparison between the beta values. Although most values are close, some outliers can be found in the categorical variables, being sex, diabetes diagnosis, and education category. However, when comparing mean absolute errors (Table \ref{tab:HASE_comp}), these differences seem to only make little impact. 

\newpage
\subsection{Survival analysis on dementia}
\begin{figure}[!htb]
    \centering
    \subfigure[]{ 
    \centering
    \includegraphics[scale = 0.25]{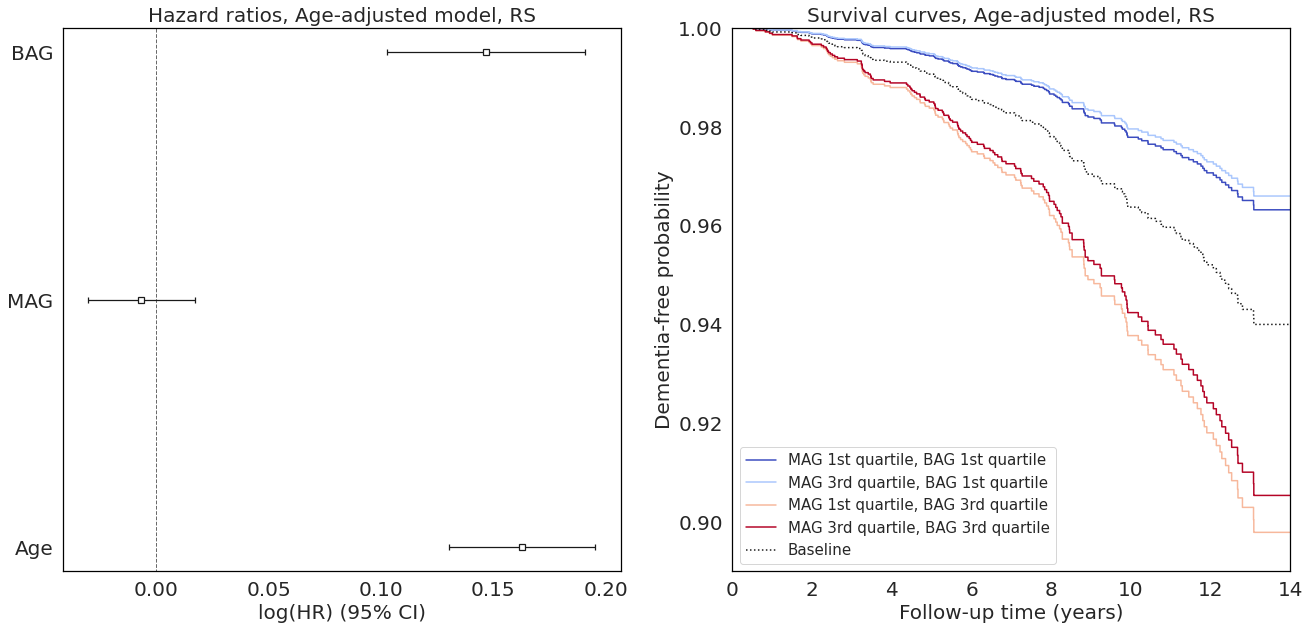}
    }
    \hfill
    \subfigure[]{
    \centering
    \includegraphics[scale = 0.25]{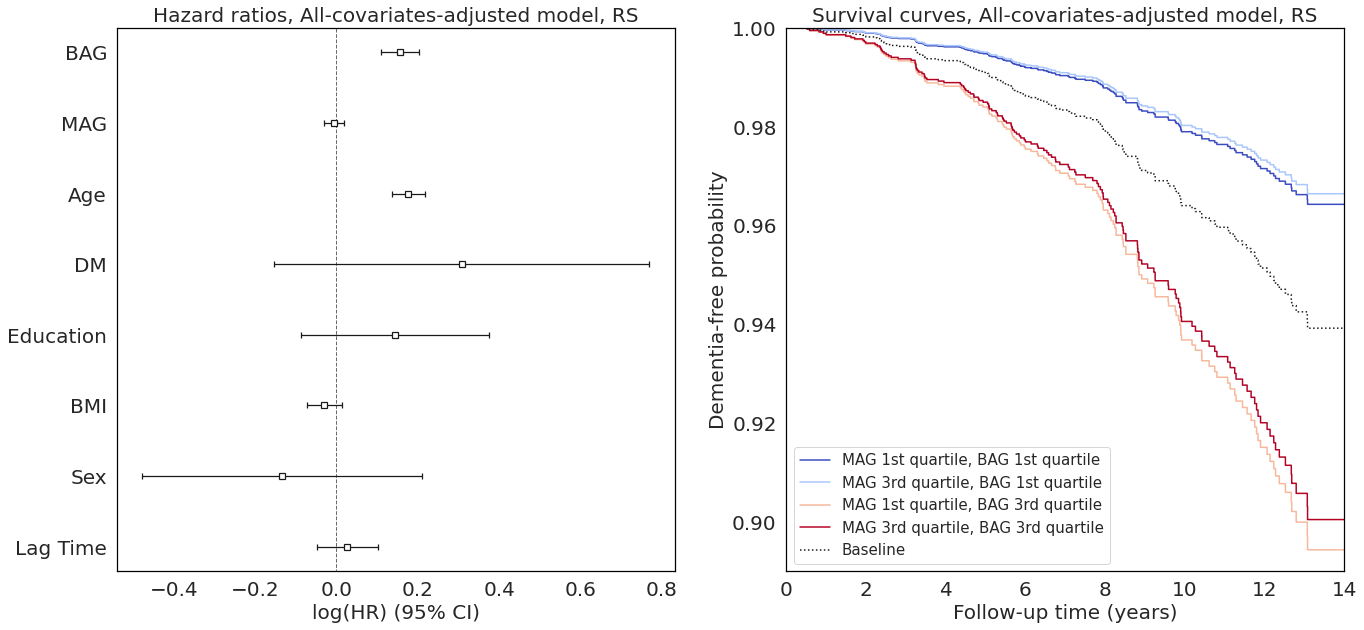}
    }
    \caption{Survival analysis for dementia prediction in RS using CPH models. On the left of (a) and (b) show the hazard ratios of the age-adjusted and all covariates-adjusted models, respectively. On the right show the survival curves of both models.}
    \label{fig:cph_dementia}
\end{figure}
\clearpage